\documentclass[pdflatex, sn-basic]{sn-jnl}% Basic Springer Nature Reference Style/Chemistry Reference Style
\usepackage{graphicx}%
\usepackage{multirow}%
\usepackage{amsmath,amssymb,amsfonts}%
\usepackage{amsthm}%
\usepackage{mathrsfs}%
\usepackage[title]{appendix}%
\usepackage{xcolor}%
\usepackage{textcomp}%
\usepackage{manyfoot}%
\usepackage{booktabs}%
\usepackage{algorithm}%
\usepackage{algorithmicx}%
\usepackage{algpseudocode}%
\usepackage{listings}%
\theoremstyle{thmstyleone}%
\theoremstyle{thmstyletwo}%
\theoremstyle{thmstylethree}%
\newcommand{\ion}[2]{\mbox{#1\,\textsc{#2}}}
\usepackage{aas_macros}
\begin{document}
\title[The Formation of Solar Prominences]{The Formation of Solar Prominences: Plasma Origin and Mechanisms}
\author*[1,2,3]{\fnm{Yuhao} \sur{Zhou}}\email{yuhaozhou@nju.edu.cn}
\affil[1]{\orgdiv{School of Astronomy and Space Science}, \orgname{Nanjing University}, \orgaddress{\city{Nanjing}, \postcode{210023}, \country{People's Republic of China}}}
\affil[2]{\orgdiv{Key Laboratory for Modern Astronomy and Astrophysics (Nanjing University)}, \orgname{Ministry of Education}, \orgaddress{\city{Nanjing}, \postcode{210023}, \country{People's Republic of China}}}
\affil[3]{\orgdiv{Centre for mathematical Plasma Astrophysics (CmPA)}, \orgname{KU Leuven}, \orgaddress{\street{Celestijnenlaan 200B}, \city{Leuven}, \postcode{B-3001}, \country{Belgium}}}
\abstract{Solar prominences, or solar filaments, are cool and dense plasma structures in the hot solar corona, whose formation mechanisms have remained a fundamental challenge in solar physics. 
This review provides a comprehensive overview of the current theoretical, numerical, and observational understanding of prominence formation, with a focus on the origin of the dense plasma component. 
We begin by summarizing the magnetic field configurations that enable prominence support, followed by a classification of four representative plasma formation mechanisms: injection, levitation, evaporation--condensation, and in-situ condensation. 
Each mechanism is analyzed in terms of its physical basis, numerical realizations, and observational diagnostics. 
A central focus is placed on the evaporation--condensation scenario, which has seen significant development over the past decade through numerical simulations. 
We also discuss recent progress in modeling in-situ condensation triggered by magnetic reconnection and levitation dynamics. 
Throughout, we emphasize the complementary nature of different mechanisms and their potential coexistence in forming and maintaining prominence mass. 
Observational constraints and recent high-resolution data are reviewed to assess the physical plausibility of each mechanism. 
We conclude by highlighting open questions and future directions in connecting multi-scale physical processes to the observed diversity of prominence behaviors.}
\maketitle

\section{Introduction}\label{sec1}
Solar prominences, often referred to as solar filaments, are dense and relatively cool plasma structures suspended in the hot solar corona \citep[reviewed by][]{hira1985, zirk1989, tand1995, mack2010, pare2014, engv2015, gibs2018, chen2020}. 
They are among the most common features observed on the quiet Sun. 
Typically, prominences exhibit temperatures around $10^4$~K, which is about two orders of magnitude lower than that of the surrounding coronal environment ($\sim 10^6$~K). 
In contrast, their densities are approximately two orders of magnitude higher than that of the ambient corona, allowing for a rough pressure balance to be maintained.

\begin{figure}[h]
\centering
\includegraphics[width=0.9\textwidth]{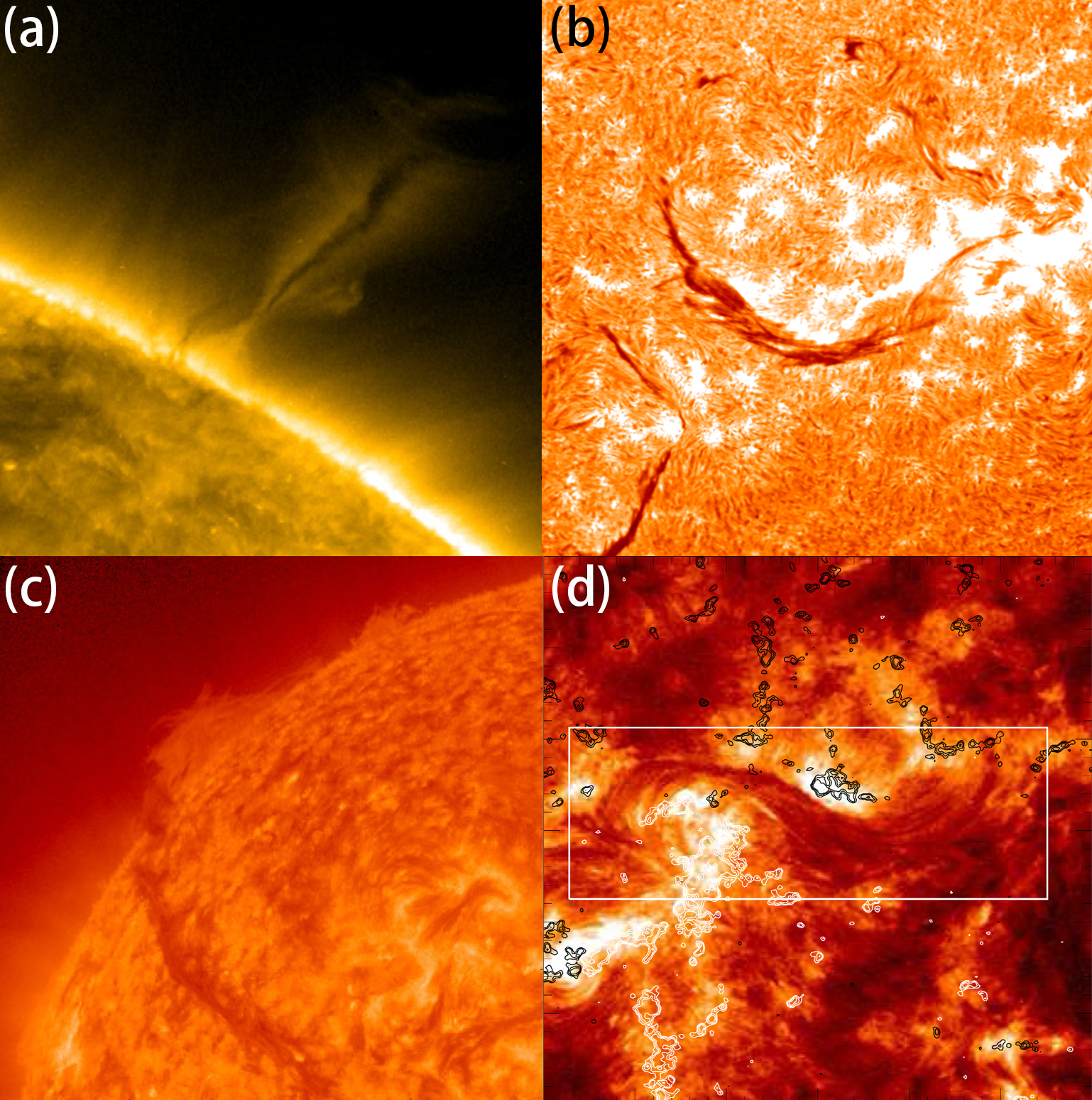}
\caption{
(a) A prominence observed by the Atmospheric Imaging Assembly \citep[AIA,][]{leme2012} onboard the Solar Dynamics Observatory \citep[\textit{SDO},][]{pesn2012} in the 171~\AA\ channel, adapted from \citet{suy2012}. 
(b) A filament imaged in H$\alpha$ by the Chinese H$\alpha$ Solar Explorer \citep[CHASE,][]{lic2022}, courtesy of the CHASE team. 
(c) Co-spatial prominence and filament structures observed by \textit{SDO}/AIA in the 304~\AA\ channel, adapted from \citet{pare2014}. 
(d) A representative example illustrating the typical alignment of a filament with the polarity inversion line (PIL), adapted from \citet{yanx2013}. The background shows an \textit{SDO}/AIA 304~\AA\ image, overlaid with white and black magnetic field contours outlining positive and negative photospheric polarities from the associated HMI magnetogram. 
}
\label{fig1}
\end{figure}

When observed at the solar limb, these structures appear bright as they protrude beyond the solar disk and are thus termed \textit{prominences}, see Fig.~\ref{fig1}(a). 
Conversely, when viewed against the solar disk, their higher densities lead to absorption of background radiation, causing them to appear as dark, elongated features known as \textit{filaments}, see Fig.~\ref{fig1}(b). 
In the early history of solar physics, prominences and filaments were thought to be distinct phenomena. 
It was not until the early 20th century, with advances in observational techniques, that it became widely accepted that they are in fact the same physical structures viewed from different angles, as shown in Fig.~\ref{fig1}(c). 
In this paper, the terms \textit{prominence} and \textit{filament} will be used interchangeably.

When compared with magnetograms (maps of the solar surface magnetic field, typically representing the line-of-sight or radial component), one finds that the main axis, or \textit{spine}, of a solar filament typically lies along the polarity inversion line (PIL), i.e. the narrow boundary separating regions of opposite magnetic polarity. 
{To illustrate this characteristic configuration, Fig.~\ref{fig1}(d) presents an example in which the filament is seen in EUV absorption and overlaid with contours marking positive and negative photospheric magnetic polarities. 
The filament clearly occupies the interface between the two opposite-polarity regions, demonstrating its typical placement above the PIL.}

Quiescent filaments{— found outside active regions and comprising intermediate and polar-crown types —} tend to be long—often extending tens to hundreds of megameters—and can remain stable for hours, days, or even weeks. 
In contrast, active-region filaments are generally shorter in both length and lifetime. 
{Intermediate filaments occur at lower latitudes outside active regions, whereas polar-crown filaments form at higher latitudes in generally weaker-field regions. 
AR filaments typically lie lower in the corona, while polar-crown filaments are often higher-lying; however, the altitude can vary with the local magnetic configuration. 
These two quiescent subtypes display distinct morphologies and may well be formed through different mechanisms.}

Since their discovery, prominences have often been likened to clouds in Earth's atmosphere. 
While this analogy captures certain visual similarities, the physical mechanisms responsible for their suspension differ fundamentally. 
Clouds on Earth are composed primarily of water droplets, whose total mass constitutes only a small fraction of the surrounding air mass. 
As a result, their overall density does not differ significantly from that of the ambient atmosphere, allowing them to remain suspended for long periods simply through convective air motion.
In contrast, solar filaments possess densities that are several orders of magnitude higher than that of the surrounding corona. 
Their levitation against gravity requires the presence of strong Lorentz forces provided by the solar coronal magnetic field, which is more structured and intense (relative to Earth's magnetic field) and capable of supporting such dense plasma against solar gravity.

The study of solar filaments involves a wide range of physical processes in solar physics, making it a particularly complex topic. 
First, the formation of filaments is intimately linked to the evolution of the solar magnetic field as well as the thermodynamics of coronal plasma. 
Second, filaments {often} exhibit oscillations and support various magnetohydrodynamic (MHD) wave modes throughout their lifetimes \citep{arre2018}. 
Third, {a large fraction of filaments eventually erupt, although some are observed to dissipate gradually without major eruptions,} expelling mass and energy into interplanetary space. 
Observations show that large-scale filament eruptions are frequently associated with coronal mass ejections (CMEs) and solar flares, making them important drivers of space weather disturbances. 
As such, a better understanding of the physical processes surrounding filaments is also of practical importance for improving space weather forecasting capabilities.

This review focuses on the understanding of filament formation. 
Research in this area has spanned several decades, and centers on two fundamental questions: (1) {what is the magnetic structure that supports the filament against gravity, and how can it be reliably constrained by observations, before we can even address its origin,} and (2) where does the dense filament plasma come from? 
The first question also lies at the heart of space weather studies, as it relates to the buildup of magnetic energy in the corona. 
The second is likely connected to one of the most long-standing problems in solar physics—the coronal heating problem.

\begin{figure}[h]
\centering
\includegraphics[width=0.9\textwidth]{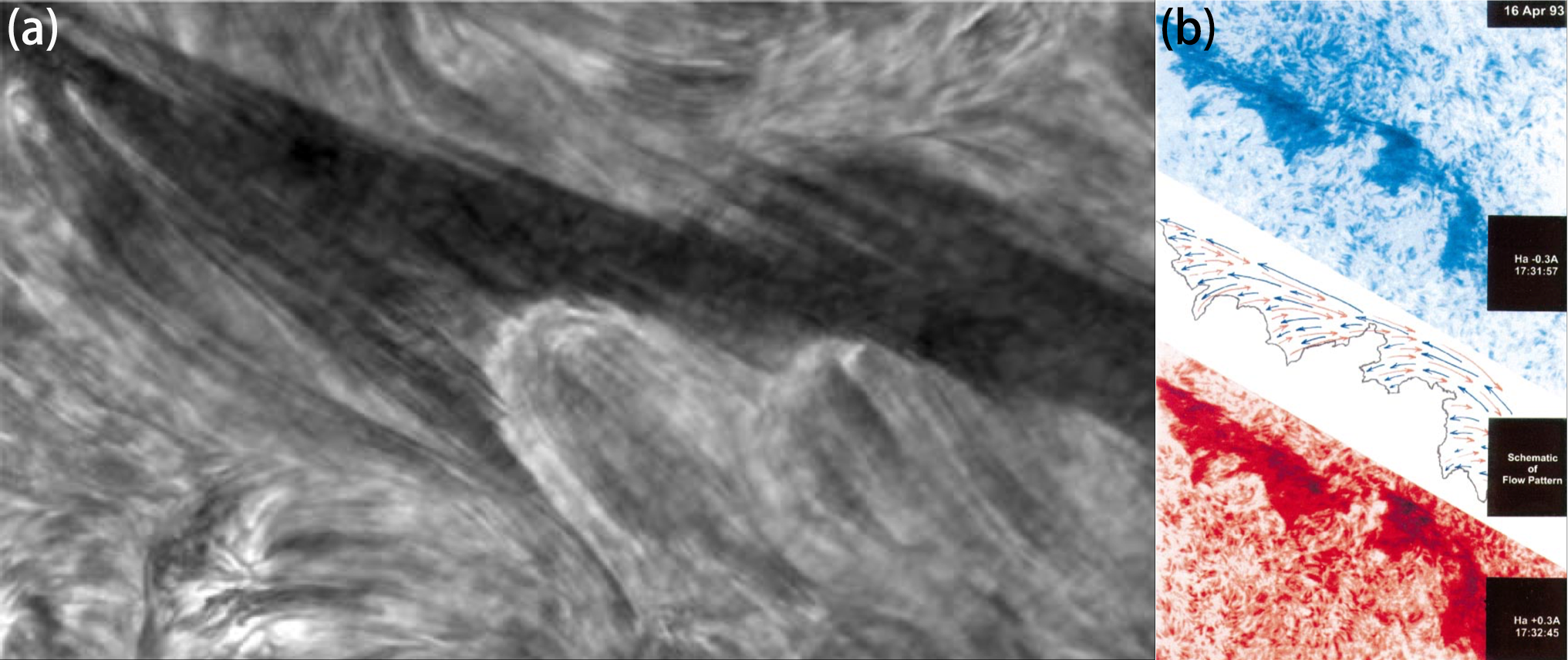}
\caption{(a) Fine-scale filament threads observed by \citet{liny2005} using SST, revealing the highly structured nature of prominence material; 
(b) The first observational evidence of counterstreaming flows along filament threads, reported by \citet{zirk1998}, highlighting dynamic plasma motions within the prominence spine.}
\label{fig2}
\end{figure}

In recent years, advances in observational capabilities—enabled by instruments such as the Swedish Solar Telescope \citep[SST,][]{scha2003}, {SDO}, Goode Solar Telescope
\citep[GST, previously NST, ][]{caow2010}, GREGOR \citep{schm2012}, the Interface Region Imaging Spectrograph \citep[IRIS,][]{depo2014}, Solar Orbiter \citep[SO or SolO,][]{mull2020}, Parker Solar Probe \citep[PSP,][]{foxn2016}, the Daniel K. Inouye Solar Telescope \citep[DKIST,][]{rimm2020}, as well as Chinese missions like New Vacuum Solar Telescope\citep[NVST,][]{liuz2014}, CHASE, and the Advanced Space-Based Solar Observatory \citep[ASO-S,][]{ganw2023}—have provided high-cadence, high-resolution, multi-wavelength data of solar prominences. 
These observations have revealed that the internal structure of filaments is far more complex than previously assumed in early models. 
In particular, high-resolution imaging shows that prominences are composed of numerous fine-scale structures, commonly referred to as \textit{threads} \citep[as shown in Fig.~\ref{fig2}(a),][]{liny2005}. 
These threads exhibit significant convective motions, oscillations, and counterstreaming flows \cite[as shown in Fig.~\ref{fig2}(b),][]{zirk1998, liny2003}. 
Moreover, their orientation, spatial scales, and relationship with the underlying magnetic topology have become active areas of research. 
Such phenomena suggest that prominence formation may not be a one-time process, but rather a dynamically sustained structure resulting from the continuous interplay of multi-scale and multi-mechanism processes.

This review aims to summarize recent progress in understanding the formation mechanisms of solar prominences, with a particular focus on how the cool, dense plasma is accumulated. 
We adopt a structure centered around three main aspects: (1) the development of theoretical models, (2) insights and validations provided by numerical simulations, and (3) observational constraints that challenge or refine existing paradigms. 
Throughout, we place these developments in the context of ongoing debates and current research frontiers.
{Complementary perspectives can be found in recent review efforts that emphasize either numerical modeling \citep{liak2025} or the broader multiphase context of prominences and rain \citep{kepp2025}}.

The structure of the paper is as follows. 
In Section~\ref{sec2}, we briefly review the magnetic field configurations associated with prominences and the general formation process of their supporting magnetic structures. 
Section~\ref{sec3} provides an overview of the theoretical mechanisms proposed for the origin of the cool plasma that constitutes the prominence body. 
In Section~\ref{sec4}, we discuss recent observational results that either support or challenge these theories. 
Section~\ref{sec5} highlights new insiƒghts from numerical simulations that contribute to our understanding of the underlying physical processes. 
Finally, in Section~\ref{sec6}, we summarize the key findings and offer a perspective on open questions and future research directions.
\newpage

\section{Magnetic Support and Field Configurations of Filaments}\label{sec2}
As previously noted, the formation of solar filaments or prominences fundamentally relies on the existence of magnetic field configurations capable of stably supporting the cool, dense plasma against gravity. 
Since the density of prominence material typically exceeds that of the surrounding coronal plasma by one to two orders of magnitude, the associated gravitational force is significantly stronger. 
In the absence of magnetic support, the material would quickly descend and dissipate. 
{In addition, the magnetic field plays a crucial role in thermally isolating the cool plasma: it strongly suppresses cross-field heat conduction, thereby allowing cold, dense structures to survive for long periods in the hot corona.} 
Therefore, understanding how such magnetic configurations form and evolve is the primary challenge in prominence formation theories, and it has been a central focus of early investigations.

\subsection{Static Magnetic Support Models}\label{sec21}
Early theoretical efforts focused on static equilibrium analyses within prescribed magnetic field structures.
These models typically assume that the prominence plasma is already present and investigate whether the forces acting within a given configuration can sustain its levitation.

Due to its high density, prominence material experiences a stronger downward gravitational force {compared to the surrounding coronal plasma}.
In magnetohydrostatic (MHS) models, this must be counteracted by an upward Lorentz force, primarily provided by magnetic tension forces (i.e., the term $\frac{1}{\mu_0} (\mathbf{B} \cdot \nabla) \mathbf{B}$, where $\mathbf{B}$ is the magnetic field vector and $\mu_0$ is the magnetic permeability). 
This implies that the supporting magnetic field lines must adopt a concave-upward configuration in regions where plasma is stably suspended.
{Furthermore, the magnetic field effectively shields cool plasma from cross-field thermal conduction, which is a crucial factor that allows the prominence plasma to remain stable in the corona for extended periods.}

\begin{figure}[h]
\centering
\includegraphics[width=0.9\textwidth]{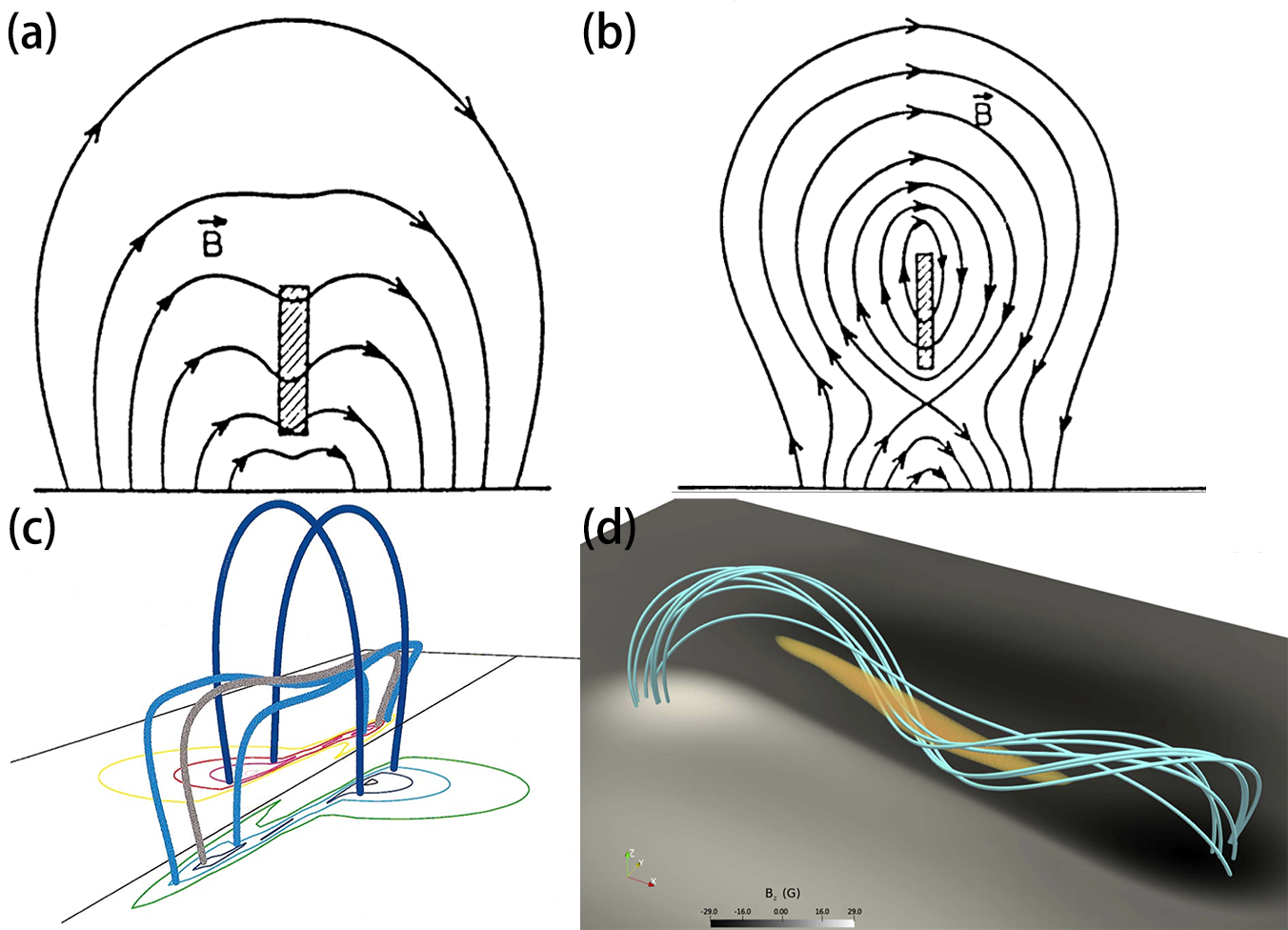}
\caption{Representative magnetic configurations that support solar prominences. 
(a) 2D schematic of the KS model, showing horizontal magnetic field lines forming a magnetic dip above the polarity inversion line (adapted from \citet{anze1985}); 
(b) 2D schematic of the KR model, in which a suspended flux rope is separated from the underlying bipolar field by electric currents (adapted from \citet{anze1985}); 
(c) 3D sheared arcade configuration (adapted from \citet{devo2000}); 
(d) 3D magnetic flux rope structure carrying a filament (adapted from \citet{zhou2018}).}
\label{fig3}
\end{figure}

Two canonical magnetic configurations have emerged from such models:

\textbf{Kippenhahn–Schlüter model} \citep{kipp1957}, commonly referred to as the KS model: This model describes nearly horizontal, arch-like magnetic field lines that bend downward near the polarity inversion line (PIL) to form magnetic dips, as illustrated in the two-dimensional (2D) schematic in Fig.~\ref{fig3}(a). 
Prominence plasma is "trapped" in these dips and is supported against gravity by magnetic tension. 
{While originally formulated in two dimensions, the KS model can also be interpreted as the limiting case of a sheared magnetic arcade; in three dimensions it naturally incorporates a guiding field component along the PIL, as illustrated by the 3D arcade configuration in Fig.~\ref{fig3}(c).}

\textbf{Kuperus–Raadu model} \citep{kupe1974}, or the KR model: This model proposes that the prominence is supported by a suspended, twisted magnetic flux rope located above the photospheric PIL, as shown in the 2D schematic in Fig.~\ref{fig3}(b). 
The flux rope is magnetically separated from the underlying bipolar field by a system of electric currents and exhibits a magnetic field direction that is opposite to that of the ambient coronal field. 
{Its three-dimensional analogue is a magnetic flux rope, illustrated in Fig.~\ref{fig3}(d). 
Statistical studies further show that the great majority of eruptive quiescent prominences are associated with flux-rope configurations \citep[e.g.,][]{ouya2017}.}

{It is worth noting that, although it is sometimes assumed that arcade-like fields (KS) correspond to normal polarity and flux-rope fields (KR) to inverse polarity, this association is not unique. 
Both arcades and flux ropes can occur in either normal- or inverse-polarity configurations \citep[e.g.,][]{gibs2018}. 
For instance, the sheared arcade shown in Fig.~\ref{fig3}(c) provides an explicit example of an inverse-polarity arcade \citep{devo2000}.}

Both models emphasize the balance of forces under static conditions and are commonly used to investigate the stability of prominence structures. 
However, they share an important assumption: the magnetic structure that supports the prominence is already in place. 
This naturally leads to a deeper question—how are such supporting magnetic configurations formed during the evolution of the solar magnetic field at the surface?

\subsection{Formation and Evolution of Supporting Magnetic Structures}\label{sec22}
In contrast to static models, recent studies have increasingly emphasized the dynamic evolution and spontaneous formation of magnetic configurations that support prominences. 
In particular, with the growing integration of numerical simulations and observations, it has become evident that the magnetic structures required for prominence support can naturally emerge from the evolution of the solar photospheric magnetic field.

A key early contribution by \citet{vanb1989} proposed that the cancellation of magnetic flux at the photosphere can lead to the formation of a twisted magnetic flux rope above the PIL, thereby self-consistently producing a magnetic topology with dips capable of supporting cool plasma. 
Subsequently, \citet{choe1992} employed 2D MHD simulations to show that shear-driven magnetic reconnection can also help to lead to the formation of such supporting structures, highlighting the importance of dynamic evolution pathways. 
These studies laid the foundation for the development of more comprehensive magnetic structure evolution models.

Building on these early pathways, many studies have suggested that the magnetic field supporting solar prominences undergoes a three-phase evolutionary process \citep{vanb2000, mart2001, mack2006, mack2010, wang2015}, which summarizes the progression from surface magnetic activity to the development of a stable supporting configuration:
\begin{enumerate}
    \item \textbf{Flux emergence and convergence:} newly emerged bipolar magnetic regions rise, expand, and converge toward the PIL, injecting free magnetic energy into the system \citep[see, for example][]{cheu2014};
    \item \textbf{Surface shear and flux cancellation:} shearing motions and flux cancellation at the photosphere increase the non-potentiality of the overlying magnetic field. {At this stage, the magnetic configuration can gradually evolve toward a pre-eruptive state, which may take the form of either a sheared arcade or a flux rope.}
    \item \textbf{Magnetic reconnection and restructuring:} under strong shear conditions, {tether-cutting reconnection \citep{moor1992, amar1996} can convert sheared field into a twisted flux rope with embedded dips, while breakout reconnection \citep{anti1999a} mainly removes overlying flux and facilitates the rise of the filament channel by forming a flare current sheet.}
\end{enumerate}

{In this way, the framework accommodates both flux-rope and sheared-arcade scenarios, with different reconnection processes playing distinct roles in the restructuring of the magnetic field.}
A common feature across all these formation pathways is that the magnetic structures supporting prominences are themselves products of dynamic evolution and should not be treated as static background fields.
However, the present review does not focus in detail on the formation of the magnetic structures themselves. 
In the following discussion of condensation and mass accumulation mechanisms, we generally assume the prior existence of a supporting magnetic configuration. 
For comprehensive reviews on the formation and evolution of such magnetic structures, the reader is referred to \citet{mack2010} and \citet{gibs2018} and references therein.

\section{Overview of Mass Accumulation Mechanisms}\label{sec3}
In addition to the existence of a suitable magnetic configuration, the formation of solar prominences requires addressing a second fundamental question: how does the cool, dense plasma accumulate within the magnetic structure? 
Given that the temperature of prominence plasma ($\sim 10^4$~K) is two orders of magnitude lower than the surrounding corona ($\sim 10^6$~K), the formation of this material is not merely a matter of plasma transport, but is deeply coupled to thermal processes. 
This section provides an overview of the current theoretical models from the perspective of different mass accumulation mechanisms.

One long-standing debate in solar physics—dating back to the mid-20th century—is whether the cool prominence material originates from the lower solar atmosphere (e.g., chromosphere or photosphere), or condenses in situ from coronal plasma. 
\citet{bruz1972} reviewed this as an unresolved issue at the time. 
A key early observational effort by \citet{niko1971} during a total solar eclipse analyzed some emission lines of metallic species in quiescent prominences. 
They found that these metallic lines were significantly enhanced relative to the background corona and spatially well-confined, suggesting a different plasma composition that may not originate from the corona. 
Although the theory of first ionization potential (FIP) bias had not yet been developed, their work represents one of the earliest attempts to infer prominence plasma origin based on spectroscopic diagnostics.

The FIP bias theory, in brief, posits that during the transport of material from the lower solar atmosphere to the corona, elements with low FIP (e.g., Fe, Mg) are preferentially transported over high-FIP elements (e.g., Ne, O), leading to compositional differences between coronal and photospheric plasmas \citep[see, for example][]{lami2015}. 
By comparing the abundance ratios of such elements—e.g., Fe/O or Mg/Ne—one can infer whether the prominence material originates directly from the lower atmosphere or is formed through in situ coronal condensation.

Using data from the Skylab mission, \citet{widi1986} measured the Ne/Mg and O/Mg ratios in an eruptive prominence and found that they were closer to photospheric values than to typical coronal FIP-enhanced levels, thus supporting a lower atmospheric origin.
\citet{spic1998} conducted a systematic analysis of seven quiescent prominences and, based on \ion{Ne}{VI} / \ion{Mg}{VI} line ratios, found that the derived Ne/Mg values were consistently higher than typical coronal values ($\sim$0.7) and closer to the photospheric level ($\sim$3.4), further reinforcing the idea that the prominence material may have been injected from the lower atmosphere.

These early studies laid the groundwork for subsequent observational efforts. 
In recent works, \citet{song2017} used in situ plasma measurements from the Advanced Composition Explorer (ACE) spacecraft in a magnetic cloud following a prominence eruption, and found that the Fe/O and Mg/O ratios were again consistent with photospheric abundances. 
Similarly, \citet{pare2019} used UV spectra from the Solar Ultraviolet Measurements of Emitted Radiation \citep[SUMER,][]{wilh1995} onboard the Solar and Heliospheric {Observatory \citep[SOHO,][]{domi1995} combined} with differential emission measure (DEM) inversions\footnote{DEM is a method for inferring the temperature distribution of plasma along the line of sight from its observed emission. For methodological details, see, for example \citet{test2012}, \citet{guen2012a}, \citet{guen2012b}, \citet{plow2013}, and \citet{cheu2015}.} and confirmed that two quiescent prominences exhibited nearly photospheric compositions with negligible FIP bias.
These multi-scale and multi-method observational results consistently suggest that, at least for some prominence structures, the plasma is more likely to be transported directly from the lower atmosphere, rather than condensed locally from coronal plasma.

It should be noted, however, that while FIP bias diagnostics have been widely adopted to infer plasma origin, they are themselves subject to significant observational and theoretical uncertainties. 
On the one hand, abundance determinations depend strongly on spectral modeling, atomic data, and assumptions about the temperature structure \citep{pare2019}. 
On the other hand, the physical mechanisms responsible for FIP bias—such as wave propagation, magnetic topology, and collisional processes—are still not fully understood, and there is no universally accepted model for their spatial extent and efficiency. 
Therefore, conclusions based on FIP diagnostics should be cross-validated with independent methods whenever possible.

Based on the presumed origin of prominence plasma, existing models can be broadly categorized into two classes. 
In one class, the plasma is thought to originate from the lower atmosphere—such as the chromosphere or photosphere—and is transported into the magnetic structure via processes such as jets, siphon flows, or evaporation-induced upflows.
In the other class, the plasma is believed to condense locally within the corona through {\textit{thermal non-equilibrium} (TNE) and \textit{thermal instability} (TI), which will be further discussed later.} 
It is important to note that since the corona itself is continuously supplied by mass from the lower atmosphere, the distinction here refers specifically to \textit{whether the prominence material is transported over short timescales from the lower layers, or forms gradually from already coronal material}.

\subsection{Injection Model}\label{sec31}
\citet{anc1988} was the first to systematically investigate the feasibility of the "injection mechanism" through numerical simulations. 
Unlike magnetohydrostatic (MHS) models, which assume the prior existence of magnetic dips capable of supporting cool plasma, injection models reverse the paradigm: they propose that the injection of plasma from the lower solar atmosphere can induce deformation of the magnetic field under the influence of gravity, thereby forming magnetic dips self-consistently. 
This concept addressed the previously underdeveloped aspect of plasma origin in prominence formation theories and laid the theoretical foundation for injection-based pathways.

Injection models posit that the cool prominence plasma can be directly "launched" from the lower atmosphere into magnetic dips. 
{In the classical picture, this injection is often envisioned as the direct transport of chromospheric cool plasma. 
However, this is an idealized scenario: energetic events in the low atmosphere can hardly convert their energy purely into kinetic form without significantly heating the plasma. 
Similar to jet phenomena \citep[e.g.][]{yoko1995}, the injected material is likely to consist of both cool and hot components, and in some cases the hot component may even dominate.} 
The formation process may involve the following steps:
\begin{itemize}
    \item In active regions, impulsive events such as small-scale reconnection or chromospheric jets trigger upward ejection of plasma;
    \item The plasma is guided along arched or curved magnetic field lines and injected into regions where magnetic dips either pre-exist or are dynamically formed due to the downward pressure of the injected mass;
    \item {Subsequently, the injected material can settle into a stable configuration within the magnetic structure (see Section~\ref{sec33} for details).}
\end{itemize}

This mechanism has been supported by a growing number of observational evidence, particularly from high-resolution instruments such as IRIS, SDO, and ground-based telescopes like GST or NVST. 
Numerous observations have revealed the presence of "cool flows" being directly transported from the lower atmosphere into filament channels. 
{More recently, several studies have also reported cases of ``hot injection,'' where plasma heated by magnetic reconnection is injected into filament channels and subsequently cools to form stable structures.} 
These signatures confirm key aspects of the injection process, especially in dynamically evolving systems.
{Representative examples include the TRACE observations by \citet{chae2003}, 
which first established a causal link between chromospheric jets and filament spine formation, 
and the complete formation sequence captured by \citet{wang2018}, 
where recurrent jet activity gradually built up a stable filament channel.} 

In particular, the injection model can naturally account for the observed high level of dynamism in prominences, where cool plasma appears to be in continuous circulation and dynamic equilibrium with the chromosphere. 
Phenomena such as counterstreaming flows can be self-consistently explained without invoking additional mechanisms: a simple pressure imbalance at the footpoints of individual magnetic threads is sufficient to drive such flows and maintain the overall mass balance.

Due to its intrinsically dynamic nature, the injection mechanism is particularly applicable to filaments that form on short timescales (typically less than tens of minutes) and are closely associated with magnetic activity. 
Such cases include active-region filaments and the rapid formation of filament spines preceding eruptions. 

\subsection{Levitation Model}\label{sec32}
\citet{rust1994} proposed this model as a new mechanism to explain the formation of prominences and filaments. 
Its core idea is that, as a magnetic flux rope rises from beneath the photosphere into the corona, it can ``scoop up'' cool material from the lower atmosphere (e.g., the chromosphere and transition region), thereby forming the initial core of the prominence structure. 
This core may subsequently grow through additional mass loading processes, such as condensation or injection.

In this model, the magnetic flux rope is initially oriented horizontally and buried below the photosphere. 
{As flux-emergence simulations show \citep[e.g.,][]{leak2014, tori2017}, the ends of the tube typically remain anchored below the surface, while the central portion rises into the corona, forming an $\Omega$-shaped loop. 
This geometry could in principle provide a channel for lifting chromospheric plasma into the upper atmosphere under magnetic tension support.
However, notice should be taken that these numerical experiments indicate that a clear dipped structure appears only in specific quadrupolar configurations \citep[e.g.,][]{tori2017}, and the outcome is sensitive to the relative orientation between the emerging tube and the overlying coronal field, indicating that the magnetic configurations suitable for levitation may not be very common.}

The model emphasizes several key features:
\begin{itemize}
    \item Prominence material is not formed through radiative cooling or condensation, but is instead directly lifted from the lower atmosphere during the upward motion of the flux rope;
    \item The scooped-up cool material becomes distributed along the lower part of the rising structure, potentially corresponding to the future filament spine;
    \item Subsequently, {the lifted plasma can accumulate within the flux rope structure, and its spatial distribution may evolve due to continued magnetic reconfiguration or additional mass supply.}
\end{itemize}

From the perspective of plasma origin, the levitation model shares similarities with injection models: both suggest that chromospheric material can directly enter the corona. 
However, the levitation model involves a more gradual dynamic evolution and a stronger role for magnetic restructuring. 
{At the same time, numerical experiments show that photospheric mass often prevents the flux rope from fully rising into the corona, or delays its ascent until reconnection produces a separate coronal flux rope (see Sect.~\ref{sec53} for details). 
In such cases, only a small amount of cool material is retained at coronal heights, as most of it drains along the rising field lines. 
This raises significant challenges for levitation as a general explanation of prominence mass supply, indicating that its role is likely limited.}

Nevertheless, the levitation scenario remains a physically plausible pathway: even if the plasma is only partially ionized, strong collisional coupling between ions and neutrals allows both components to be lifted together \citep{terr2015}. 
{Given these considerations, levitation may be relevant in some restricted circumstances—for example, during the early accumulation stage or in special magnetic topologies—while in most observed cases, additional processes such as injection or evaporation-condensation are likely required to account for the prominence mass reservoir.}

\subsection{Evaporation--Condensation Model}\label{sec33}
Under typical physical intuition, stronger heating leads to higher plasma temperatures, while weaker heating results in lower temperatures. This principle formed the basis of many prominence formation models developed in the 1970s and 1980s. 
In contrast, the evaporation--condensation mechanism operates in a regime where this intuition breaks down. 
Its core physics relies on the interplay between two key processes: \textit{thermal non-equilibrium} (TNE) and \textit{thermal instability} (TI).

TNE refers to a state in which the energy input (typically from heating at the loop footpoints) and energy losses (primarily via radiation and conduction) are not perfectly balanced over time. 
In this condition, plasma slowly accumulates and cools along coronal loops in a quasi-steady way, without immediately destabilizing the system—typically as a result of chromospheric evaporation driven by localized footpoint heating. 
TI, on the other hand, occurs when a small perturbation—such as a local increase in density—causes radiative losses to rapidly outpace heating, triggering a runaway cooling process \citep{park1953}. 
{It is important to note that the precise relationship between TNE and TI is debated. 
Some studies describe condensation as a progression from TNE to TI (e.g., \citealt{xiac2011, clae2019}), whereas others argue that the two are fundamentally distinct: TI requires a well-defined equilibrium as the starting point, whereas TNE assumes that no stable solution exists under highly footpoint-concentrated heating \citep{klim2019}. 
This distinction implies that not all TNE states necessarily lead to TI, and that rapid cooling episodes may be interpreted either as TI or as the nonlinear manifestation of TNE depending on the adopted definition.}

The theoretical basis for this process dates back to the mid-20th century. 
\citet{park1953} and \citet{fiel1965} proposed the theory of TI, demonstrated that in high-temperature, low-density plasma systems, a slight imbalance between energy input and radiative losses can destabilize the thermal equilibrium and trigger catastrophic cooling, leading to localized condensation and plasma densification—a scenario that provides a plausible physical basis for the formation of cold structures such as prominences.
Later, \citet{anti1980}, in studying post-flare coronal loop cooling, emphasized that localized density enhancements can significantly accelerate radiative cooling and trigger rapid development of TI. 
He noted that when the coronal plasma reaches densities of $\sim 10^{11}$~cm$^{-3}$, radiative losses can dominate even at coronal temperatures ($\sim 10^6$~K), potentially leading to catastrophic cooling. 
This density-driven pathway is closely aligned with the physical logic of the evaporation--condensation scenario, where evaporation precedes plasma accumulation and eventual cooling.

Since the formal introduction of the evaporation--condensation model by \citet{anti1999b}, it has become one of the most widely accepted frameworks for prominence formation. 
The typical sequence involves:
\begin{enumerate}
    \item Strong localized (possibly asymmetric) heating is applied near the chromosphere or transition region, particularly at magnetic footpoints;
    \item This concentrated heating drives evaporation of plasma into the coronal loop, particularly into concave-upward regions or magnetic dips;
    \item As plasma accumulates in the dip, its density increases and temperature gradually decreases, pushing the system into a thermally TNE state;
    \item {As the system evolves, a phase of rapid cooling sets in, which may be understood either as the onset of TI or as the nonlinear development of TNE, depending on the physical conditions and interpretation.}
    \item The condensed plasma stabilizes at chromospheric temperatures, forming a prominence or filament thread.
\end{enumerate}

This process generally assumes the presence of magnetic dips in the configuration, which facilitate long-term accumulation and growth of cool plasma threads. 
Even if the magnetic field or heating is not perfectly symmetric—which is typically the case in real solar conditions—the dip still acts as a gravitational trap that enables thread elongation. 
In cases where no dip is present, two outcomes are possible: either the accumulated cool material bends the magnetic field downward to form a dip (namely, similar with the injection model, returning to the above scenario), or the plasma drains along the field lines due to gravity. 
The latter case is more akin to \textit{coronal rain} than prominence formation—while physically related, the two phenomena differ significantly in observational appearance and spatial coherence.

In recent years, a number of numerical simulations, further reviewed in Sect.~\ref{sec5}, have enriched our understanding of this mechanism.
A series of one-dimensional (1D) models have systematically explored the effects of heating asymmetry, loop geometry, loop length, and condensation location, and have led to the development of multi-thread statistical models for simulating filament thread dynamics. 
Subsequent multi-dimensional simulations have incorporated inter-field-line heat exchange, anisotropic thermal conduction, and self-consistent formation of dips at loop tops, significantly improving both the physical realism and observational relevance of the model.

Similar to the injection model, the evaporation--condensation mechanism can also account for the observed dynamic behavior of prominences. 
Through continuous cycles of condensation and drainage, the system can maintain a quasi-steady state in which cool plasma is constantly formed, falls back, and is replenished. 
Although the maintenance of this dynamic equilibrium is less intuitively apparent than in the injection scenario, numerical simulations \citep[e.g.][]{xiac2016} have demonstrated that a self-sustained mass cycling process is indeed achievable within this framework.

It is worth noting that the evaporation--condensation and injection models share many similarities. 
Both rely on localized energetic events at chromospheric or transition region footpoints to transport plasma into the corona, where it subsequently accumulates in magnetic dips and forms prominence structures. 
The key difference lies in how energy is converted and transported. 
In the injection model, the energy of small-scale events is primarily converted into kinetic energy, akin to outflows driven by impulsive magnetic reconnection. 
In contrast, the evaporation--condensation model channels most of this energy into heating, causing chromospheric plasma to evaporate upward. 
This process resembles a more gradual reconnection scenario, where energy is deposited predominantly as thermal energy rather than kinetic energy. 
Recent studies have also attempted to unify this mechanism with injection models. 
For example, \citet{huan2021, huan2025} showed that impulsive localized heating applied at different heights can separately trigger evaporation--condensation or injection processes, suggesting that the two mechanisms are not mutually exclusive but may instead represent different responses of the solar atmosphere to energy input. 
{From this perspective, injection and evaporation--condensation should not be regarded as strictly disjoint: 
injection is typically associated with more impulsive, kinetic-dominated processes, whereas evaporation--condensation reflects a more gradual, heating-dominated pathway, with intermediate or mixed cases naturally expected in the dynamic solar atmosphere.}

{Nevertheless, despite many similarities among the models, current observations, particularly those of quiescent filaments, suggest that the evaporation--condensation scenario has certain advantages. 
This is because the injection model requires a continuous supply of cool plasma from the chromosphere into the corona during formation, whereas the evaporation--condensation paradigm necessarily produces hot plasma along the same field line between the chromosphere and the forming prominence. 
This implies that signatures such as funnel prominences without chromospheric connections, or filament threads of finite length that do not extend to the surface \citep[e.g.,][]{liny2005, hein2006}, are naturally explained by evaporation--condensation. 
Such differences provide valuable diagnostics for distinguishing between competing formation mechanisms.}

In summary, the evaporation--condensation mechanism is firmly grounded in physical theory and supported by increasingly sophisticated simulations. 
However, key open questions remain—particularly regarding the nature of the heating processes which require further theoretical modeling and observational validation.

\subsection{In-situ Condensation Models}\label{sec34}
The evaporation--condensation mechanism highlights how sensitive the local thermal balance can be under certain coronal conditions. 
In a typical coronal environment, thermal conduction along magnetic field lines generally maintains stability; however, localized inhomogeneities in heating or radiative losses can still lead to departures from equilibrium.
Once the system evolves into a state of TNE, further increases in density or changes in heating can lead to a phase of rapid cooling and plasma condensation, due to the onset of TI or the nonlinear development of TNE.

Importantly, this {sensitivity of the thermal equilibrium} implies that strong evaporation driven by highly localized heating is not a prerequisite for initiating condensation.
Instead, a wide variety of disturbances—whether in the form of impulsive heating, magnetic reconfiguration, wave-induced compression, or loss of conductive flux—may suffice to disrupt the delicate energy balance in the corona and induce in-situ condensation. 
This broadens the range of physical conditions under which prominence plasma can form and motivates the development of alternative, non-evaporative condensation models.

Direct observational evidence for the coexistence of cool plasma within the hot solar corona was provided by \citet{lero1972}. 
By using long-exposure coronagraphic imaging, he {identified localized weak H$\alpha$ emission features above the solar limb, 
demonstrating that small-scale, low-density cool plasma elements could be embedded in coronal structures and remain detectable against the bright coronal background.} 
Although no definitive formation mechanism was proposed at the time, this discovery challenged the prevailing assumption of an isothermal corona and provided {early evidence} for the coexistence of cold material in the coronal environment.

A major theoretical advancement followed with the work of \citet{levi1977}, who investigated the thermal evolution of coronal loop systems following the abrupt cessation of energy input.
They found that such loops could undergo rapid radiative cooling and drain a substantial fraction of their mass—up to 70\%—toward the lower atmosphere. 
The inferred cooling timescales were significantly shorter than expected from radiation losses alone, indicating the presence of a catastrophic TI triggered by the loss of heating. 
Although spatial resolution at the time precluded the direct detection of condensations, their analysis clearly demonstrated that, even in the absence of chromospheric mass supply, the corona itself could spontaneously generate cool, dense plasma via internal thermal imbalance. 
This work established the physical foundation for the modern paradigm of in-situ condensation.

Together, the observational findings of \citet{lero1972} and the theoretical modeling of \citet{levi1977} laid the groundwork for contemporary studies of coronal rain and prominence formation, both of which rely on the fundamental idea that TNE within the corona can lead to the spontaneous generation of cool plasma condensations.

Subsequent studies further explored this cooling scenario from both observational and theoretical perspectives. 
While some of these efforts gradually evolved into the evaporation--condensation framework, others retained the notion that condensations could originate directly within the corona, laying the conceptual foundation for modern in-situ condensation models.
In-situ condensation models posit that even in the absence of strong chromospheric input, the corona can still experience sufficient thermal imbalance to generate prominence material. 
This marks a fundamental departure from both the evaporation--condensation and injection models, which depend on mass input from the lower atmosphere, whereas in-situ condensation models propose that prominence plasma originates self-consistently from coronal material alone. 
{However, an important caveat was already raised by \citet{pike1971}, who argued that the quiet corona does not contain enough mass to supply even one or two prominences, implying that the entire coronal mass would have to condense. 
This mass-budget problem was one of the main motivations for the development of the evaporation--condensation model. 
Later estimates, however, suggest that typical filament masses are small compared to the low-coronal reservoir \citep[see, e.g.,][]{chen2020}, so the severity of the depletion remains debated. 
We therefore emphasize that the adequacy of the coronal mass reservoir is still an open question. 
Given this uncertainty, in-situ condensation may be more relevant in post-eruptive environments where the ambient coronal density is enhanced, as suggested by observations of funnel and spider prominences \citep[e.g.,][]{liuw2012, berg2012, pana2019}.}

A prominent example {within the framework of in-situ condensation} is the \textit{levitation--reconnection} model, also named as \textit{reconnection--condensation} model, in which magnetic reconnection forms a flux rope that lifts overlying coronal plasma into a closed magnetic structure, creating favorable conditions for condensation. 
In this scenario, the reconfiguration of magnetic topology serves as the primary trigger for cooling and mass accumulation. 

This contrasts with the earlier levitation model, where cool plasma is directly transported into the corona to form the initial prominence. 
In the levitation--reconnection scenario, only a portion of cool material is uplifted, and while it may not immediately constitute a prominence, the increased density lowers the threshold for radiative instability, thereby facilitating subsequent in-situ condensation.

This model was proposed and numerically demonstrated by \citet{kane2015, kane2017}, and is characterized by several key features:
\begin{itemize}
    \item The condensation process occurs entirely in the coronal environment, without requiring chromospheric evaporation or injection of cool material;
    \item The onset of condensation depends on both magnetic topology (e.g., loop length) and TI criteria;
    \item It is particularly suited to explaining quiescent prominences that form in the absence of strong low-atmosphere activity.
\end{itemize}

As noted earlier, the coronal thermal balance {can be sensitive under certain conditions.
Apart from levitation, various types of perturbations may lead to spontaneous condensation.} 
Alternative in-situ scenarios have been proposed and demonstrated by numerical simulations.
{For example, post-flare condensation was already noted in early studies (e.g., \citealt{anti1980}) and has since been further explored by modern simulations,} including post-flare loop-top condensation \citep[e.g.,][]{ruan2021} and post-eruption cooling-driven condensation \citep[e.g.][]{sens2024}. 
{A common feature of these post-eruptive models is that the eruption elevates the local coronal density, thereby not only providing sufficient perturbations but also supplying additional mass to enable in-situ condensation.} 
{Although most current simulations reproduce coronal rain rather than long-lived prominences, the underlying thermodynamic mechanisms are closely related, and it is plausible that similar processes could lead to prominence formation in post-eruptive environments. 
Nevertheless, further dedicated simulations are required to firmly establish this connection.}

In-situ condensation models thus expand the theoretical boundary of prominence mass origin by emphasizing the coupling between magnetic evolution and thermal transport. 
However, similar to the levitation model, in-situ condensation may face challenges in explaining how prominences—particularly quiescent ones—can persist in a highly dynamic state over long durations. 
Although its basic mechanism is conceptually similar to evaporation--condensation, in-situ condensation lacks a stable mass reservoir, such as the chromosphere, from which plasma can be continuously supplied. 

Although some simulations have demonstrated the possibility of establishing a self-sustained mass cycle relying solely on coronal plasma \citep[e.g.,][]{kane2018}, it remains uncertain whether such a mechanism is robust under realistic solar conditions—particularly given the continual mass loss driven by prominence drainage and the limited mass reservoir in the low corona.
Relying solely on coronal material, it is unclear whether a sustainable mass cycle can be established, especially given the continual mass loss caused by prominence flows. 
Without an external source to replenish this loss, maintaining the overall prominence mass may be difficult. 
This limitation suggests that additional mechanisms—such as chromospheric injection or evaporative replenishment—may be necessary to support long-term prominence survival in a dynamically evolving coronal environment.

\subsection{Coexistence of Multiple Formation Mechanisms}\label{sec35}
Although the injection, levitation, evaporation--condensation, and in-situ condensation models each present self-consistent theoretical frameworks and have been validated through numerical simulations, a growing number of observational and theoretical studies suggest that these mechanisms are not mutually exclusive. 
Instead, they may coexist and interact across different spatial and temporal scales during the formation and evolution of prominences.

For example, the main body of a prominence may be formed via evaporation--condensation, while barbs or localized enhancements could arise from episodic plasma injection. 
Moreover, different branches within the same magnetic structure may undergo distinct thermodynamic evolution paths, with some regions forming condensations via TI, and others accumulating chromospheric plasma through jet-like injection. 
This spatial and temporal hybridization implies that no single mechanism can fully account for the observed diversity of prominence structures.

Moreover, high-resolution observations from IRIS, SDO, and ground-based telescopes consistently reveal that prominences are highly dynamic structures \citep[e.g.][]{zirk1998, liny2003, dier2018}. 
Plasma within filaments exhibits counterstreaming motions and continuous mass circulation, suggesting that individual threads have short lifetimes and are constantly replenished. 

Such behavior implies that the prominence system exists in a dynamic equilibrium, rather than as a static mass confined in magnetic dips. 
However, among the models discussed, the levitation mechanism inherently lacks a self-consistent explanation for this sustained dynamism, while in-situ condensation models also face challenges in maintaining rapid mass cycling under coronal conditions. 
These limitations further underscore the need for multi-mechanism frameworks that can account for both the formation and long-term maintenance of prominence plasma under continuously evolving conditions.

It should also be noted that, {in numerical models, TNE and TI can often be triggered by modest localized perturbations such as variations in heating or density.} 
However, whether such processes can robustly occur in the real corona—and result in long-lived, spatially coherent prominences—remains an open question. 

In summary, the study of prominence formation is gradually shifting from viewing the proposed mechanisms as competing alternatives to recognizing them as complementary pathways. 
{At the same time, different prominence types may favor different dominant mechanisms—for example, active-region filaments may be more consistent with injection, while quiescent prominences often show signatures more compatible with evaporation--condensation.} 
Future progress will require a more integrated approach that considers both their coexistence and their possible type-dependent dominance, along with robust observational diagnostics. 
Bridging theoretical modeling with multi-scale, multi-wavelength observations will be key to advancing our understanding of this complex and dynamically evolving solar phenomenon.

\section{Observational Perspectives on Prominence Formation Mechanisms}\label{sec4}
Although multiple theoretical models have been proposed to explain prominence formation, identifying which of these mechanisms actually occur—and under what conditions—remains a major observational challenge. 
This difficulty arises in part from limited spatial and temporal resolution, as well as from the restricted diagnostic capability for multi-thermal plasma. 
In many cases, the mere appearance of cool material is insufficient to unambiguously determine the underlying formation process.

Based on the four representative formation scenarios introduced in Section~\ref{sec2}, we propose the following observational classification framework to interpret recent results:

\begin{itemize}
    \item \textbf{In-situ condensation}: Cool material is generated through TNE and TI processes within the corona, without requiring external injection or significant plasma transport from lower atmosphere. 
    Observationally, this is characterized by the apparent ``spontaneous'' appearance of condensations or periodic formation cycles, without associated upflows or localized heating signatures.

    \item \textbf{Evaporation--condensation}: Prominence plasma originates from chromospheric evaporation, which subsequently cools and condenses in coronal magnetic dips. 
    In contrast to in-situ condensation, this mechanism emphasizes the upward transport of hot plasma prior to cooling. 
    Observationally, it manifests as evaporation flows or siphon-like upflows with a measurable temperature evolution from coronal to prominence regimes.

    \item \textbf{Injection}: Cool plasma is directly ejected from the lower atmosphere into magnetic dips, without undergoing prior heating or evaporation. 
    This scenario is supported by observations of upward-moving cool jets, H$\alpha$ surges, or the gradual accumulation of cool threads in filament channels.

    \item \textbf{Levitation}: The formation of cool material occurs as it is lifted from the lower atmosphere by a rising magnetic flux rope. 
    This process does not necessarily involve strong plasma flows. 
    Observationally, it is characterized by the bulk rise of magnetic structures, with low-atmosphere material being transported upward passively along the field lines.
\end{itemize}

Since in-situ condensation and evaporation--condensation share many observational characteristics—both typically presenting as cooling and condensation within coronal loops—they are often difficult to distinguish based solely on imaging or spectral diagnostics. 
Therefore, in the following sections, we treat these two mechanisms jointly under the category of ``condensation-driven'' formation. 
We begin by discussing observational evidence for the injection scenario, which tends to exhibit more direct kinematic signatures, and subsequently turn to the condensation-driven and levitation scenarios, supported by a range of recent studies.

\subsection{Injection Model}
The injection model proposes that cool chromospheric or even photospheric plasma is dynamically transported along magnetic field lines into coronal magnetic dips, thereby forming prominence structures. 
Over the past two decades, this mechanism has gained substantial observational support.

\subsubsection{Early evidence: causal links between jets, injection, and structure formation}
Early direct observational evidence for the injection model was provided by \citet{chae2003}, who used TRACE imaging to demonstrate for the first time that chromospheric jets, originating from magnetic flux cancellation sites, can be guided along magnetic field lines into an inverse-S-shaped filament spine. 
This work established a causal link between jet activity, mass supply, and filament formation. 
Building on this, \citet{liuy2005} conducted a broader survey using multi-wavelength observations and found that chromospheric jets are frequently observed prior to prominence formation, with a statistical correlation to subsequent spine development. 
More recently, \citet{zoup2016} reported high-resolution H$\alpha$ observations of small-scale fibrils with upward velocities of 5–-10 km/s, originating from chromospheric brightenings associated with photospheric flux cancellation. 
These jet-like flows, likely driven by magnetic reconnection, persisted for tens of minutes and reached altitudes consistent with the expected magnetic dips. 
Collectively, these studies demonstrate both the physical plausibility and the observational ubiquity of injection-driven mass loading processes in active region prominences.

\subsubsection{Full formation sequence: from initially unfilled corona to prominence spine}
\begin{figure}[h]
\centering
\includegraphics[width=0.9\textwidth]{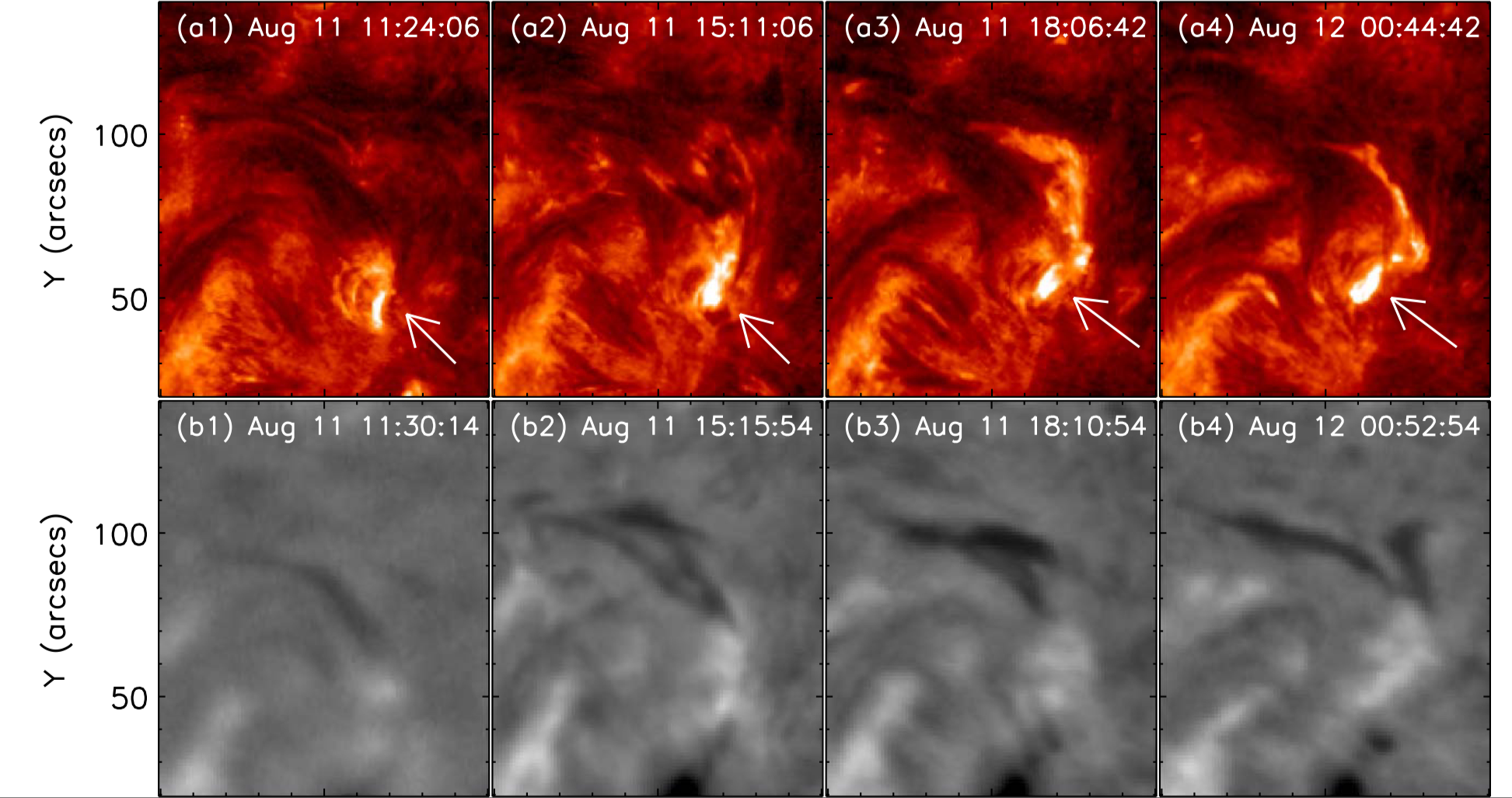}
\caption{
Example of filament formation driven by recurrent chromospheric jets, adapted from \citet{wang2018}.  
Panels (a1)--(a4) show a sequence of SDO/AIA 304~\AA{} images capturing collimated jet-like ejections from the western footpoint of a forming filament.  
The white arrows highlight jet events that contribute to plasma injection along the filament channel.  
Panels (b1)--(b4) present the corresponding GONG H$\alpha$ observations taken shortly after each jet episode, showing progressive accumulation of cool material within the filament spine.  
This event illustrates a typical observational scenario in which localized, repetitive jet activity supplies mass to a growing filament structure.
}
\label{fig4}
\end{figure}

\citet{wang2018} presented one of the first complete observations of a prominence formation from an {initially unfilled coronal volume, where no pre-existing cool plasma structures were visible,} to a well-developed spine, as shown in Fig.~\ref{fig4}. 
They reported successive injections of chromospheric plasma guided by magnetic field lines into the coronal environment, which gradually accumulated to form a stable filament channel.
In a follow-up study, \citet{wang2019} combined observations from SDO and NVST to estimate the total injected mass, finding a value of approximately $1.6 \times 10^{15}$~g—comparable to the estimated mass of the fully developed prominence. 
The injected material exhibited a multi-thermal character, with both hot components visible in EUV and cooler flows observable only in H$\alpha$.

Extending this picture, \citet{guoy2021} tracked the full sequence of injection and prominence spine formation. 
They found that jet-like flows consistently preceded the appearance of the filament spine and {reached heights consistent with the expected locations of magnetic dips}, {which is consistent with} the magnetic guidance of injected material. 
The presence of downward-draining flows further suggested a feedback process between mass loading and magnetic topology. 
Collectively, these studies provide strong observational support for the injection model and represent some of the few datasets that capture the full prominence formation process {starting from an initially unfilled coronal volume with no pre-existing, observable cool-plasma structures.}

\subsubsection{Applicability in quiet-Sun regions: limitations and edge cases}
While the injection mechanism is well supported by observations in active regions, its applicability in quiet-Sun environments remains under debate. 
\citet{zoup2019} reported a jet event originating from a weak-field region, where the initial upward velocity reached 20--30~km/s. 
However, due to the shallow magnetic inclination, the plasma was only lifted to a height of approximately 10~Mm—substantially lower than the typical heights ($>$25~Mm) associated with quiescent prominences. 
The authors argued that the limited kinetic energy and unfavorable magnetic geometry in such regions may constrain the efficiency of the injection process. 
In contrast, \citet{sunx2023} observed sustained injection of cool plasma and subsequent formation of a small-scale prominence at the periphery of an active region. 
This suggests that the injection mechanism may still operate under intermediate magnetic conditions, provided that sufficient inclination and connectivity are present. 
These findings underscore the need for further investigation into the magnetic and energetic thresholds that govern the spatial applicability of the injection scenario across different solar environments.

\subsubsection{Complex mechanisms and hybrid processes: injection--condensation coupling}
With the increasing resolution of solar observations, it has become evident that plasma injection is frequently accompanied by additional processes such as localized heating and in-situ condensation. 
\citet{lih2023} utilized combined CHASE and SDO observations to investigate a merging site of two filament threads and revealed persistent jet activity originating from transition-region temperatures ($\sim$$10^5$~K). 
These jets were closely associated with plasma accumulation and bidirectional flows, suggesting a hybrid scenario where transition-region jets may serve as triggers for condensation processes. 
Similarly, \citet{yang2024} documented a flare-driven event in which chromospheric jets formed a complete funnel-like prominence spine within 30 minutes, demonstrating the potential efficiency of the injection mechanism under strongly energized conditions.

In contrast, \citet{yang2023} reported an injection scenario where hot plasma (6–8~MK) generated by magnetic reconnection was transported into a newly formed filament channel and later cooled to form a stable U-shaped filament, {consistent with the broadened picture of the injection model that allows heated plasma to be injected rather than only pre-cooled chromospheric material.} 
Similarly, \citet{zhan2024} observed thread-by-thread reconnection between low-lying filaments that triggered hot plasma flows into a high-lying magnetic channel, which subsequently became visible as a coherent filament structure during the plasma’s gradual cooling. 
{These ``hot injection'' cases therefore enrich the phenomenology of the injection model, illustrating that reconnection-driven injections can supply both cool and hot components, with the latter cooling in situ to form stable filament structures.}

More recently, \citet{lih2025} presented the first full-sequence observation of cool plasma rising quasi-periodically along arcade-like magnetic field lines and gradually assembling into a coherent filament spine. 
These upward-moving threads had temperatures below $10^4$~K and velocities of 5–12~km/s, yet displayed remarkable spatial coherence and temporal stability. 
The authors argued that this behavior was likely driven by persistent, small-scale magnetic reconnection rather than explosive dynamics, highlighting the diversity of physical conditions under which injection and condensation may jointly operate.

\subsubsection{Summary and outstanding questions}
Injection mechanisms are now supported by a substantial body of observational evidence in the context of active-region prominences. 
From the initial identification of upward-moving cool jets to full reconstructions of structure formation and analysis of combined mechanisms, a coherent and multifaceted evidence base has emerged.

Nevertheless, several unresolved questions remain:
\begin{itemize}
    \item Is the injected material pre-cooled before ejection, or does it cool during its ascent? Current diagnostics lack the resolution to clearly distinguish between these scenarios;
    \item Most observed injection events occur in active regions; whether this mechanism applies to quiescent prominences is still uncertain;
    \item The causal relationship between mass injection and magnetic structure formation requires further kinematic inversion and statistical analysis to be conclusively established.
\end{itemize}

Future progress will require a combination of high-resolution magnetic field extrapolations, spectroscopic diagnostics, and multi-channel imaging to systematically characterize the triggering, transport, and cooling of injected plasma, and to assess the role of this mechanism across different prominence classes.

\subsection{Condensation Model}
\subsubsection{Condensation Signatures}
Direct observational evidence for rapid cooling and in-situ condensation in the corona has been reported for decades, particularly in the context of coronal rain. 
Although most early detections involved coronal rain rather than prominences, the underlying cooling mechanism—driven by local thermal imbalance—is now widely considered common to both phenomena, differing mainly in magnetic configuration and scale. 
As such, many studies of coronal rain remain relevant for evaluating condensation as a pathway for filament material.

{Early indications came from multi-wavelength imaging and spectroscopy, where coronal loops were observed to cool from coronal to transition-region temperatures and to produce high-speed downflows \citep[e.g.,][]{kjel1998, shim2002}. 
These results confirmed that cool plasma structures could form \textit{in situ} in the corona through spontaneous radiative or conductive cooling, without requiring chromospheric mass input. 
Later studies with Hinode/SOT Ca~II~H further established coronal rain as an indirect but robust tracer of localized thermal instability, showing condensations forming near loop tops and draining along both legs \citep{anto2010, anto2012}. 
A significant breakthrough came with the series of works by \citet{from2015,from2017,from2018}, who linked long-period EUV pulsations to thermal non-equilibrium cycles driven by asymmetric footpoint heating, culminating in condensation and plasma fallback.}

{More direct evidence for in-situ condensation leading to prominence formation has also been obtained with \textit{SDO}/AIA.} 
\citet{berg2012} reported a spectacular event in which cool dense plasma appeared within a coronal cavity without clear chromospheric injection, strongly supporting condensation occurring in the corona itself. 
Similarly, \citet{liuw2012} presented the first detection of prominence condensation by \textit{SDO}, where cool plasma formed in the corona following a flaring event. 
These studies provide compelling observational examples that prominence mass can condense in situ in coronal structures. 
More recently, \citet{huan2023} observed the formation of a new filament following the radiative cooling of a hot coronal loop, with no clear evidence of chromospheric evaporation. 
{These observations suggest} that coronal condensation may independently contribute to filament mass loading in certain cases. 
{However, at the same time, as discussed in Sect.~\ref{sec34}, the ultimate adequacy of the coronal mass reservoir for in-situ condensation remains debated. 
Future observations that can provide quantitative constraints on the pre-formation coronal density, or at least establish systematic links between in-situ condensation events and post-CME environments, would offer powerful diagnostics for resolving this mass-source issue.}

\subsubsection{Evaporation Signatures and Triggering Mechanisms}
Beyond direct condensation, {it is important to note that most of the observational evidence discussed so far has captured only the cooling and condensation phases, with little or no direct detection of the preceding evaporation flows—largely due to instrumental limitations. 
Only in recent years have observations begun to capture the full evaporation--condensation pathway. 
Representative studies again first emerged in the context of coronal rain. 
For example, \citet{kohu2019} analyzed coordinated SDO/AIA and IRIS data of NOAA AR~12468, where localized reconnection produced impulsive heating in the upper leg of a coronal loop, followed by evaporation, in-situ condensation, and subsequent coronal rain drainage. 
Similarly, \citet{from2020} combined multi-wavelength SDO/AIA data with high-resolution SST imaging to confirm a complete evaporation--condensation--drainage cycle. 
Their analysis showed that systematic heating and density increase preceded the condensations, providing multi-scale, multi-thermal validation of TNE-driven cycles as a mechanism for generating cool plasma in the corona.}

\begin{figure}[h]
\centering
\includegraphics[width=0.9\textwidth]{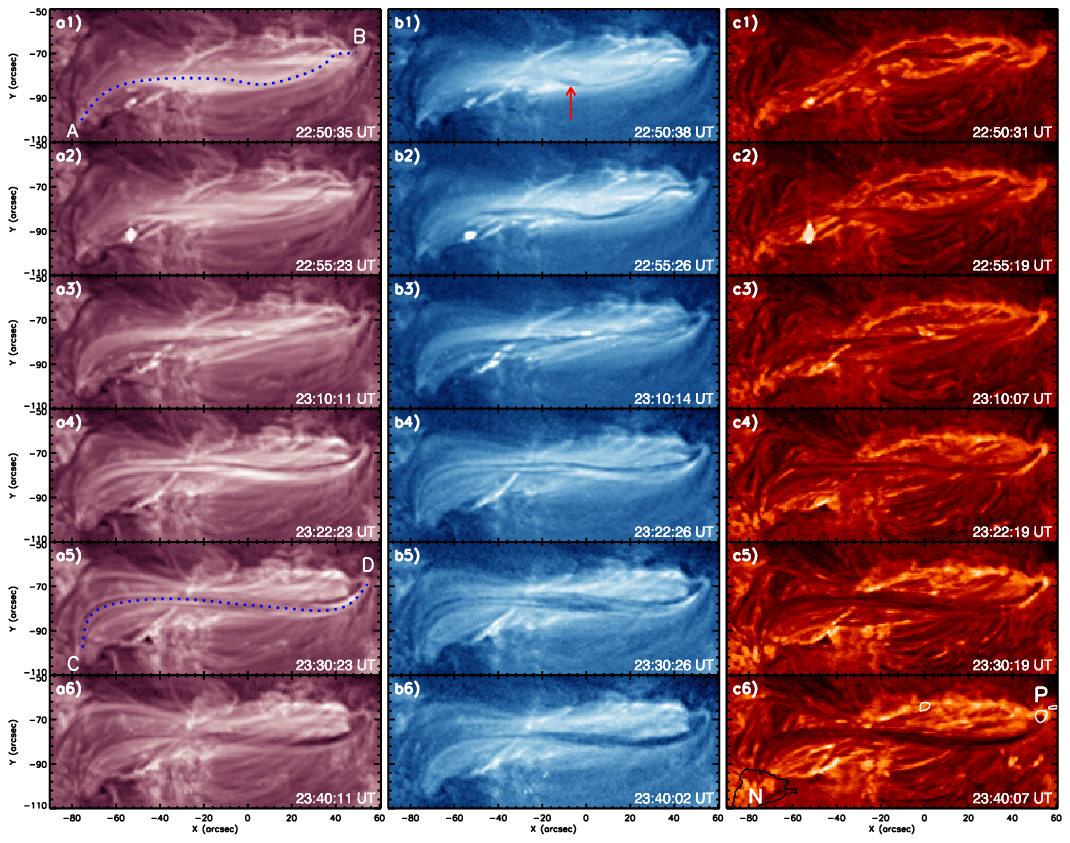}
\caption{
Example of filament formation through coronal condensation, adapted from \citet{yang2021}.  
Panels (a1)--(a6), (b1)--(b6), and (c1)--(c6) show the evolution of the event in SDO/AIA 211~\AA{}, 335~\AA{}, and 304~\AA{} channels, respectively.  
The red arrow in panel (b1) marks the onset of localized condensation in the corona, which subsequently develops into a coherent filament structure as plasma cools and accumulates.  
This event represents a well-resolved observational case of the evaporation--condensation process, in which thermally unstable plasma condenses in situ without apparent chromospheric injection.  
The presence of evaporation is supported by Hinode/EIS spectroscopic measurements showing significant blueshifts in high-temperature coronal lines (Fe~\textsc{xv}, Fe~\textsc{xvi}) at the flare footpoints, indicative of upward chromospheric evaporation flows.
}
\label{fig5}
\end{figure}

{Building on these developments, }\citet{lil2021} and \citet{yang2021} provided the first observational evidence that the evaporation--condensation mechanism also operates in filament formation, marking a major step beyond coronal rain toward prominence-relevant contexts.
In both studies, magnetic reconnection triggered heating at the chromospheric footpoints, leading to chromospheric evaporation that supplied hot plasma into newly formed coronal magnetic structures. 
Subsequent cooling and in-situ condensation then formed distinct H$\alpha$ filaments. 
Particularly, \citet{yang2021} presented a spectroscopically confirmed sequence of explosive evaporation followed by in-situ condensation, establishing a complete observational chain linking magnetic reconnection to filament formation, as shown in Fig.~\ref{fig5}. 

These observations share several key characteristics:
\begin{itemize}
  \item The injected plasma typically originates in the transition region or low corona, with initial temperatures of $10^5$–$10^6$\,K;
  \item Triggering mechanisms include localized reconnection or transient heating at loop footpoints;
  \item Condensation occurs preferentially at {or near} magnetic dips, consistent with observed prominence morphology;
  \item Thermal evolution proceeds from hot injection to cool accumulation over observable timescales.
\end{itemize}

Collectively, this body of work demonstrates that the evaporation--condensation mechanism is not merely a theoretical construct, but a physically observable process that manifests under specific magnetic and thermal conditions.  
While most prominently identified in quiescent and intermediate filaments, it may also operate in active regions when appropriate constraints are met.  

Nevertheless, current observational evidence for evaporation--condensation in prominences tends to be associated with relatively energetic events, such as flares or strong localized heating episodes.  
It remains uncertain whether similar processes operate under the more modest energy conditions typical of most quiescent filaments.  
This limitation likely reflects the current instrumental resolution and sensitivity. 
Future high-resolution, high-cadence observations will be essential to determine how widespread and energetically diverse this mechanism truly is in filament formation.

\subsection{Levitation Model}
The lifting or ``emergence'' model was first proposed by \citet{rust1994}, who also presented direct observational evidence from a prominence formation event on 1989 April 27. 
Multi-wavelength data revealed a rising magnetic flux rope (MFR) exhibiting a characteristic geometry with $\Omega$-shaped legs and a central U-shaped dip. 
Crucially, cool plasma was observed within the dip and was lifted into the corona as the structure rose, suggesting a process distinct from top-down condensation.  

They also reported the gradual weakening and disappearance of small magnetic polarities at the rope's footpoints, interpreted as reconnection accompanying the rope's emergence through the photosphere. 
Given the near-simultaneity between rope emergence and prominence formation—occurring faster than radiative cooling timescales—\citet{rust1994} proposed that pre-existing cool material was directly carried upward, rather than condensing post-emergence.

This model received observational support in the high-resolution study by \citet{deng2000}, who used data from SVST and TRACE to examine a filament formation event in AR 8331. 
Their observations revealed the upward emergence of a U-shaped magnetic structure containing cool material, bounded by expanding $\Omega$-loops. 
Associated photospheric magnetic cancellation and downflows near the footpoints (up to 20~km/s) were consistent with flux rope emergence and the ``scooping'' of material into the atmosphere. 
UV brightenings in TRACE data further indicated that the magnetic reconfiguration impacted overlying coronal structures.

Although this event closely matched the geometric and dynamic expectations of the lifting model, some researchers \cite[e.g.][]{mack2010} have classified it as part of a broader category of reconnection- or relaxation-driven ``lifting-like'' mechanisms, rather than ideal flux emergence.

\citet{lite2005} conducted high-precision vector magnetogram observations beneath active region filaments and identified concave-up magnetic field geometries at the photospheric level, consistent with the expected configuration of emerging flux ropes.  
While the study did not capture the dynamic process of plasma being lifted, the observed field topology and associated magnetic flux cancellation provide geometric support for the emergence scenario.
\citet{xuz2012} further examined the emergence of a twisted magnetic flux rope in a decaying active region using vector magnetic field measurements and multi-line imaging.  
They reported a concave-up field configuration in the photosphere, upward-directed chromospheric flows along the polarity inversion line, and the gradual appearance of a coherent filament spine in He~\textsc{i}~10830\,\AA{} images.  
While the cool material did not appear to be bodily lifted with the emerging field, it progressively accumulated during the emergence process, likely aided by the dipped topology of the rising flux rope.  
Thus, this event provides partial observational support for the lifting model, albeit favoring a scenario in which cold plasma is collected during, rather than prior to, the emergence.

In summary, while several studies provide compelling morphological or partial dynamic support for the lifting model, direct observations of cool plasma being bodily lifted into the corona remain rare. 
In active regions, strong magnetic activity and low-lying filament heights may obscure such processes, while in quiescent regions, the difficulty lies in sustaining stable emergence and preventing plasma drainage during ascent. 
Furthermore, plasma is not rigidly trapped in magnetic dips and may flow in response to local pressure imbalances, complicating the idealized ``scooping'' scenario. 
Whether this mechanism operates independently, or only in concert with others, remains an open question.

\section{Insights from Numerical Simulations}\label{sec5}
\subsection{Evaporation--Condensation Model}
Among various formation mechanisms, thermal condensation, particularly via the evaporation--condensation pathway, has been one of the most extensively studied processes in numerical models over the past few decades.  
It is important to note that in lower-dimensional models, the distinction between prominences and coronal rain is not always clear. 
In {1D simulations, which track plasma evolution along a single magnetic flux tube, condensations cannot be unambiguously classified as either prominences or coronal rain, since both simply appear as localized cooling events.} 
In {2D simulations, which typically represent cross-sectional cuts through magnetic structures, the situation is more nuanced: many cases can be reasonably interpreted as prominence formation, but alternative rain-like interpretations are not always excluded.} 
For this reason, {in the following discussion we do not differentiate between filamentary and rain-like condensations in 1D cases, while for 2D studies we focus on those works that explicitly demonstrate prominence formation.} 
In contrast, {3D simulations resolve both morphology and dynamics sufficiently to distinguish between these phenomena; as such, 3D simulations of coronal rain are not included in the present review.}

\subsubsection{Conceptual Development and Early Formulation}
Given the limited computational capabilities of early numerical studies, prominence formation was often modeled using 1D hydrodynamic simulations along magnetic flux tubes.  
This simplification, however, remains physically justified: under typical low plasma-$\beta$ conditions in the solar corona, plasma dynamics are effectively constrained to follow magnetic field lines, with negligible feedback on the field structure itself.  
A 3D cartoon illustrating this geometry is shown in Fig.~\ref{fig6}(a), where the flux tube is represented as a transparent channel and the prominence material appears as a dense core near the loop apex.

A seminal numerical application of the evaporation--condensation mechanism under coronal conditions was presented by \citet{moky1990}, who performed 1D nonlinear simulations of plasma dynamics along coronal loops.  
They demonstrated that if the heating is concentrated at the loop footpoints rather than at the apex, condensation can form near the loop top even in the absence of a pre-existing gravitational dip.  
Their work provided one of the earliest physical demonstrations of how TNE can lead to spontaneous condensation in a coronal environment.

Building upon this framework, \citet{anti1991} introduced a more counterintuitive physical insight: localized \textit{enhancement} of heating can itself trigger condensation.  
In their model, enhanced footpoint heating induces strong chromospheric evaporation, which increases the coronal density.  
As a result, radiative losses at the loop apex may exceed the local heating rate, leading to TI and eventual condensation.  
This mechanism challenges the conventional view that condensations are primarily caused by insufficient heating, and instead shows that under certain conditions, enhanced heating can also produce condensations through nonlinear evolution.

Following these works, {the} evaporation–condensation model was systematically introduced by \citet{anti1999b}, who formulated a complete physical sequence: localized heating $\rightarrow$ chromospheric evaporation $\rightarrow$ mass accumulation $\rightarrow$ {loss of thermal equilibrium and subsequent catastrophic cooling} $\rightarrow$ condensation.
Using 1D hydrodynamic simulations, they demonstrated the feasibility of this process and established the model’s conceptual framework, which has since become one of the leading paradigms for prominence formation.
In a follow-up study, \citet{anti2000} extended this model to show that thermal imbalance alone—even under constant heating—can naturally induce condensation and plasma motion within coronal loops. 
They further revealed that the system is highly sensitive to heating asymmetry, explaining why many condensations are transient and mobile.
This behavior aligns with observed counterstreaming flows and intermittent mass loading events in prominences \citep[e.g.,][]{mart1998, zirk1998}.
These insights strengthened the model’s explanatory power and highlighted the delicate heating configurations required for long-lived prominence support.

\subsubsection{From Early Foundations to Model Maturity}
Building on this theoretical foundation, \citet{karp2001, karp2003, karp2005, karp2006, karp2008} conducted a series of numerical studies to test and expand the applicability of the evaporation--condensation mechanism. 
They showed that condensations can form and migrate even in the absence of magnetic dips, as long as the heating is footpoint-concentrated and spatially asymmetric. 
Therefore, in shallow dips, condensations tend to fall back, while steeper dips promote long-term retention and mass accumulation.
This is consistent with the formation of the quiescent prominence spine.
To improve observational relevance, \citet{karp2005} incorporated magnetic field lines derived from 3D sheared-arcade models and used updated radiative loss functions, achieving condensation cycles that matched observed periods and dynamics. 
Subsequent simulations by \citet{karp2006} found that nearly horizontal magnetic channels facilitate thread-like condensations and generate transient high-speed flows (up to 100~km/s), offering explanations for counterstreaming and activation features. 
In \citet{karp2008}, impulsive heating scenarios (e.g., nanoflares) were modeled, showing that as long as heating intervals remained shorter than the radiative cooling time, condensations could still occur. 
This study also simulated DEM profiles and found strong agreement with observations in the range $\log T \approx 4.6$–$5.6$.

Concurrently, \citet{mull2003, mull2004} focused on transition-region and low-coronal conditions, exploring the model’s universality. 
Their 1D simulations showed that even under steady heating, if spatially localized at the footpoints, catastrophic cooling can occur without pre-existing dips. 
This yielded periodic evaporation--condensation cycles, with downflow speeds (up to 128~km/s) and deceleration signatures matching observed coronal rain. 
Their models also reproduced UV line emissions (e.g., O~\textsc{v}, C~\textsc{iv}) via non-equilibrium ionization, enhancing spectral comparability.

\begin{figure}[h]
\centering
\includegraphics[width=0.9\textwidth]{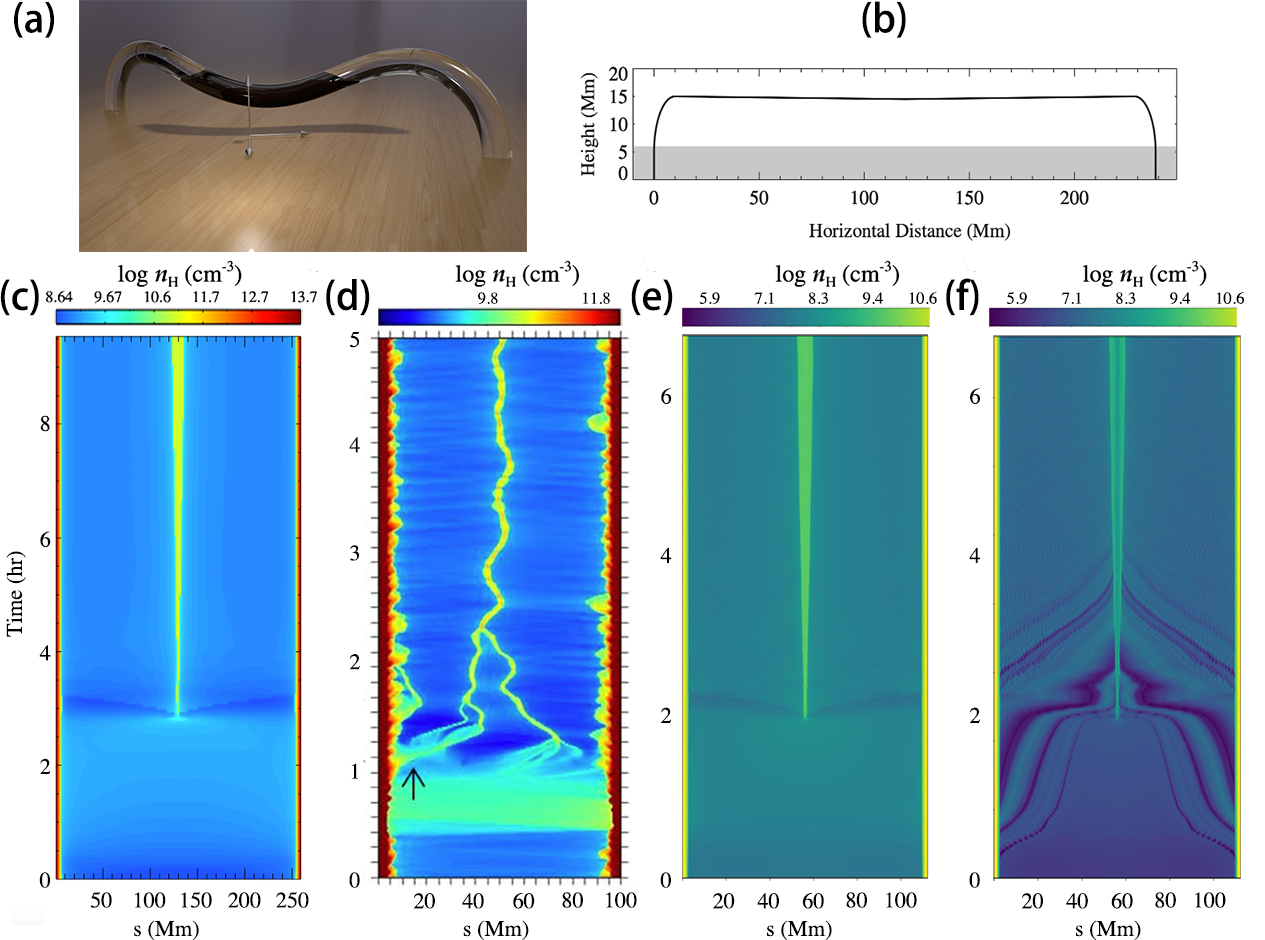}
\caption{
Recent advances in one-dimensional evaporation--condensation simulations. 
(a) 3D cartoon of a magnetic flux tube. 
The transparent structure represents a magnetic field-aligned flux tube, and the black material illustrates condensed prominence plasma.  
(b) Geometry of a typical 1D loop model, adapted from \citet{xiac2011}.  
While the specific field line shape and heating parameters may vary across studies, this basic configuration remains widely adopted.  
(c) Simulation result from \citet{xiac2011} under spatially symmetric and steady heating conditions. 
The horizontal axis represents the loop-aligned coordinate, and the vertical axis denotes time; the color scale indicates number density evolution, highlighting spontaneous condensation near the loop apex.  
(d) Simulation result from \citet{yosh2025}, demonstrating condensation formation under Alfv\'en-wave-driven heating.  
(e) and (f) Results from \citet{jerc2025}, which incorporate a two-fluid treatment. 
Panels show the number density evolution of (e) electrons and (f) neutrals, respectively, emphasizing differential condensation behavior in partially ionized plasma environments.
}
\label{fig6}
\end{figure}

Based on these works, \citet{xiac2011} systematically investigated whether finite-duration localized heating could trigger condensation.
Their 1D configuration and resulted condensation is shown in Fig.~\ref{fig6}(b) and (c), respectively.
They found that, given appropriate initial conditions, condensation could self-sustain and grow even after the localized heating ceased, driven by pressure gradients. 
This significantly expanded the model's applicability. 
Their simulations also validated Parker's isochoric TI criterion, showing that condensations form under nearly constant density with abrupt temperature drops---a result consistent with isochoric instability. 
Building upon this, \citet{zhou2014} further explored the long-term evolution of condensations after their initial formation. 
They found that even after the localized heating was removed, prominences could continue to grow due to sustained radiative losses from the surrounding corona. 
Although the growth rate diminishes over time, this self-sustained accumulation broadens the model's applicability and highlights a critical coupling point between condensation-driven processes and other mechanisms such as injection or levitation models.

Together, these studies elevated the evaporation--condensation model from a theoretical concept to a comprehensive framework capable of explaining prominence formation, dynamics, and thermal signatures under a range of magnetic and heating conditions.

\subsubsection{Refining the Model: Multiphysics and Microprocesses}
While the classical work of \citet{xiac2011} established the viability of the evaporation--condensation mechanism for prominence formation in a simplified 1D setting, more recent studies have extended the framework by incorporating increasingly realistic physical effects and modeling choices. 
These developments reflect a broader effort to bridge the gap between idealized simulations and the complex, structured nature of the solar corona.

\citet{maso2023} investigated how strong magnetic expansion near coronal null points affects loop thermodynamics. 
Compared to the symmetric geometry used by \citet{xiac2011} and many previous works, their model included significant cross-sectional asymmetry, characteristic of separatrix surfaces. 
The resulting geometry induced steady siphon flows even under uniform heating, producing flow patterns consistent with observations. While no condensation occurred, the study suggested that extreme expansion may locally suppress thermal conduction and promote conditions favorable to TI.

More recently, several studies have sought to address the limitations of artificially prescribed heating in evaporation--condensation models by introducing more realistic or physically motivated heating processes. 
\citet{kuce2024} investigated the impact of nanoflare-driven, spatially localized impulsive heating on condensation formation. 
Compared to the steady, symmetric heating used in \citet{xiac2011} and most of the previous works, their stochastic, intermittent heating schemes better reflect observational expectations of energy release in the corona. 
However, nanoflares remain a parametrized input, and their spatial distribution is not derived from first principles.
To move beyond such assumptions, \citet{yosh2025} implemented a ``fully self-consistent" Alfvén wave heating mechanism based on Elsässer variable turbulence \citep{elsa1950}, whose result is shown in Fig.~\ref{fig6}(d). 
Dissipation arises naturally from wave reflection and nonlinear interaction, eliminating the need for imposed localized heating. 
Their simulations showed that a single impulsive heating episode, embedded within a realistic turbulent background, can initiate condensation once the heating energy exceeds a critical threshold set by radiative losses and thermal conduction. 
Recently, further generalizations of wave energy transport have been proposed, such as the \textit{Q}-variable formulation introduced by \citet{vand2024}, which extends beyond the Elsässer framework to include compressive MHD wave modes. 
Early applications of \textit{Q}-driven heating are underway, and such approaches may enable even more physically complete and self-consistent coronal heating models in the near future.

In parallel, \citet{jerc2025} addressed the single-fluid limitation of earlier models by introducing a two-fluid framework for partially ionized plasma. 
Their results are shown in Fig.~\ref{fig6}(e) and (f).
This allowed explicit treatment of ion–neutral decoupling, which led to new dynamic features such as species-specific shock structures, velocity drifts, and frictional heating near the prominence–corona transition region. 
Together, these studies represent a significant step toward embedding the evaporation--condensation mechanism within a more complete and self-consistent physical framework.

\subsubsection{Insight from Incomplete Condensation to Complete Condensation}
Meanwhile, over the past decade, many numerical modelings have shifted focus from the classical formation of fully condensed cool plasma (as in prominences) to the broader framework of TNE in coronal loops. 
While incomplete condensation (IC) cycles do not directly lead to the formation of chromospheric-temperature condensations, the coronal environment established by TNE is now recognized as a favorable precursor for triggering catastrophic cooling. 
Understanding TNE, therefore, has become a central element in constraining the onset conditions of coronal condensation and mass accumulation.

The work of \citet{miki2013} explored how magnetic loop geometry and footpoint heating asymmetry influence the occurrence and nature of TNE cycles. 
Their simulations revealed that TNE can manifest in two regimes: (1) complete condensation (CC), where local plasma temperatures drop to chromospheric values ($\sim10^4$~K), and (2) incomplete condensation (IC), where the temperature remains within the coronal range ($\sim0.6$--$1.0$~MK). 
The results indicated that non-uniform cross-sectional expansion and asymmetric heating tend to favor IC-type cycles, while also allowing for sustained siphon flows that can inhibit full condensation. 
This work provided a theoretical framework to interpret the diversity of observed loop behaviors within a unified thermodynamic model.
Building upon this framework, \citet{from2017} performed 1D TNE simulations using realistic magnetic loop geometries derived from linear force-free field extrapolations of SDO/HMI data. 
Their aim was to reproduce long-period ($\sim$9 hr) EUV intensity pulsations observed by SDO/AIA. 
Adopting a double-exponential footpoint heating profile similar to that used in \citet{miki2013}, they successfully reproduced pulsation periods and temperature variations consistent with the observed event. 
The loop apex temperature remained above $0.6$~MK throughout the cycle, consistent with the IC regime and indicative of warm loop dynamics without the formation of cool condensations.

\citet{from2018} and \citet{pelo2022} extended this modeling framework through systematic parameter-space studies that varied loop geometry and heating configurations. 
These investigations collectively identified three distinct TNE regimes: static equilibrium (SE), incomplete condensation (IC), and complete condensation (CC). 
Their results emphasized that TNE cycles occur only under specific conditions where the magnetic geometry and energy deposition are suitably matched. 
Symmetric loops with balanced footpoint heating tend to favor CC-type cycles and coronal rain formation, while asymmetric setups more often produce IC-type pulsations or suppress TNE altogether. 
By exploring thousands of configurations, these studies established a comprehensive framework linking coronal loop structure and heating profiles to the thermal evolution of the plasma, showing that both EUV pulsations and condensations may naturally arise from TNE dynamics under different physical conditions.

A recent extension of TNE modeling beyond closed coronal loops was presented by \citet{scot2024}, who demonstrated that TNE cycles can also occur within open magnetic flux tubes supporting a transonic solar wind. 
Using a 1D hydrodynamic model with empirical heating, they showed that when footpoint heating exceeds a critical threshold, a localized thermal sink forms below the sonic point, leading to periodic condensation and fallback of material—effectively embedding coronal rain–like dynamics within a solar wind outflow. 
This result broadens the applicability of the evaporation--condensation framework, suggesting that TNE may also contribute to the mass and density variability in the solar wind, particularly in regions of enhanced magnetic flux concentration or at coronal hole boundaries.

In parallel to these physically motivated studies, \citet{john2019b} investigated the numerical challenges involved in simulating TNE, particularly the sensitivity of TNE behavior to spatial resolution, background heating, and the choice of numerical scheme. 
They demonstrated that insufficient resolution in the transition region (TR) can artificially suppress the development of TNE cycles, or significantly alter their period and thermodynamic signatures. 
To address this, \citet{john2019a} introduced the TRAC (Transition Region Adaptive Conduction) method to address long-standing numerical challenges in resolving the transition region (TR) in field-aligned and multidimensional coronal simulations. 
By dynamically identifying unresolved portions of the TR and broadening the thermal conduction and radiative loss profiles below an adaptive cutoff temperature, TRAC preserves the essential enthalpy exchange between the corona and chromosphere without requiring prohibitively high spatial resolution. 
In subsequent studies, \citet{iiji2021}, \citet{zhou2021}, and \citet{john2021} independently developed multidimensional extensions of the TRAC methodology within different modeling frameworks. 
Despite their independent origins, all three approaches successfully addressed the long-standing issue of suppressed enthalpy flux in under-resolved multidimensional evaporation–condensation simulations, thereby enabling the formation of condensations that were previously inhibited by numerical limitations.

\subsubsection{Multidimensional Realizations of the Evaporation--Condensation Mechanism in Magnetic Arcades}
The first 2D MHD simulations to explicitly model the evaporation–condensation process within a magnetic arcade were carried out by \citet{xiac2012} and \citet{kepp2014}.
In both studies, localized chromospheric heating was imposed at the footpoints of an arcade structure to trigger evaporation. 
The upflowing plasma accumulated near the loop tops, gradually increasing the density and reducing the temperature until TI was reached, initiating condensation. 
A key difference lies in the magnetic configuration: \citet{xiac2012} began with a dip-free arcade, allowing the weight of the condensations to bend field lines and generate dips self-consistently, whereas \citet{kepp2014} assumed a pre-existing dipped topology, which enabled longer simulation times. 
In the latter case, mass loading from condensation eventually led to magnetic reconnection, transforming part of the sheared arcade into a flux rope. 
Their simulation also captured the late-stage drainage of condensations, reminiscent of coronal rain. 
These works collectively demonstrated the viability of the evaporation--condensation mechanism in multidimensional settings.
%In Fig.~\ref{fig7}(a), we show the prominence morphology of a 3D extension of the 2D simulation by \citet{kepp2014}. 
%The setup is conceptually similar to the original 2D case, but with the arcade extended into the third dimension and with slightly non-uniform localized heating imposed along the footpoints. 
%This illustrates that the evaporation--condensation mechanism remains viable in 3D geometry and produces prominence-like structures that are more comparable to observations.

\begin{figure}[h]
\centering
\includegraphics[width=0.9\textwidth]{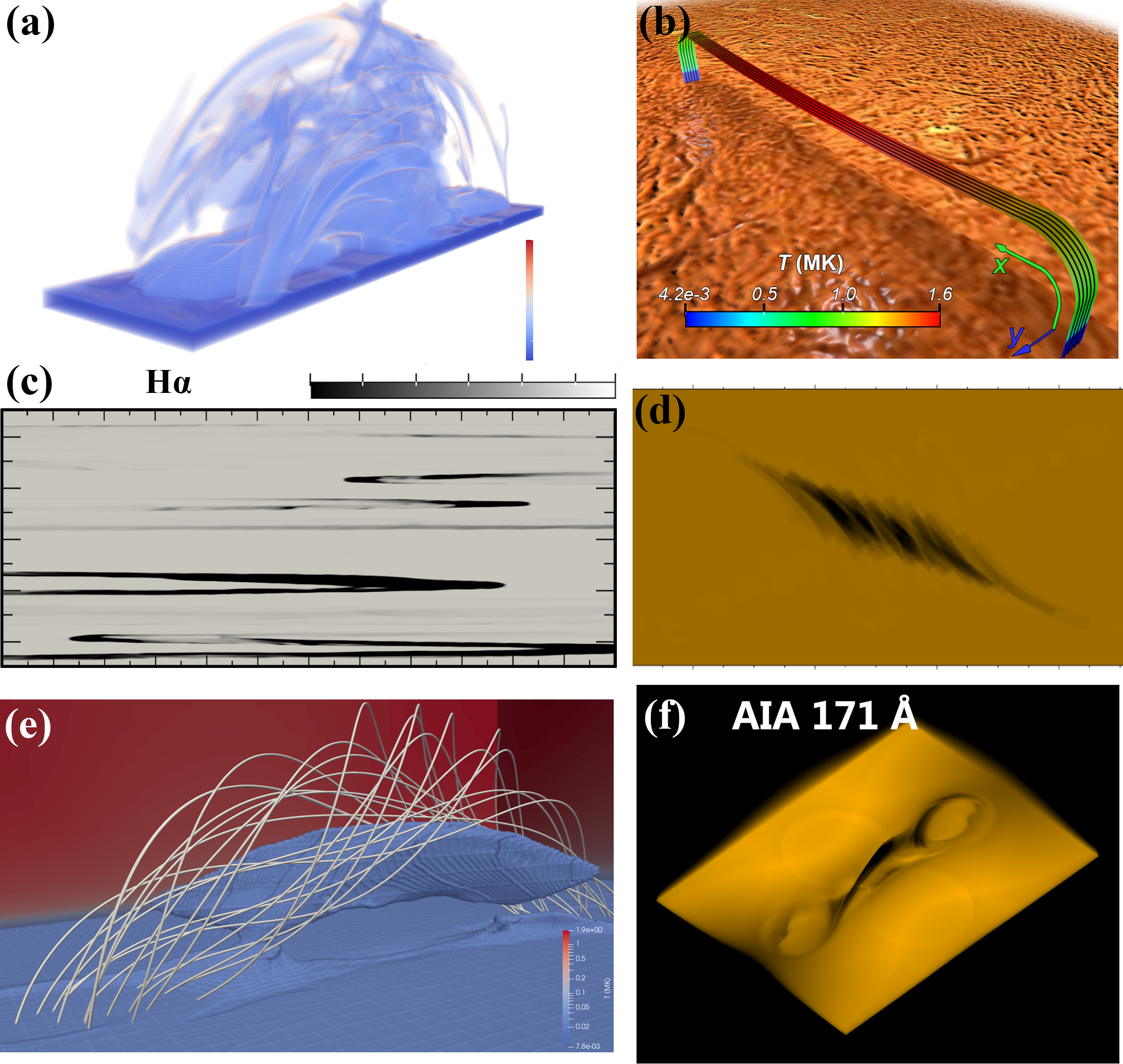}
\caption{
Recent high-resolution and three-dimensional simulations of the evaporation--condensation mechanism.  
(a) A 3D version of the 2D simulation by \citet{kepp2014}, showing the spatial complexity of condensations within a three-dimensional arcade structure.  
Temperature contours indicate the distribution of cool, condensed plasma embedded in the magnetic field.  
\textit{From the author's unpublished simulations.}
(b) Three-dimensional magnetic configuration inferred from the 2D MHD simulation of \citet{zhou2020}, illustrating slightly dipped field lines and prominence-supporting structures.  
(c) Synthetic H$\alpha$ image based on the same simulation, revealing the fine thread-like structures.  
(d) EUV 193~\AA{} synthetic emission map from the pseudo-3D simulation of \citet{guoj2022}.  
(e) Magnetic field topology and cool plasma contours from the 3D ffHD simulation of \citet{zhou2025}, capturing volumetric prominence formation.  
(f) Corresponding EUV 171~\AA{} synthetic image derived from the same simulation.
}
\label{fig7}
\end{figure}

In Fig.~\ref{fig7}(a), we show the prominence morphology resulting from a 3D extension of the 2D simulation by \citet{kepp2014}. 
The setup is conceptually similar to the original 2D case, but the arcade is now extended into the third dimension and the heating at the footpoints is prescribed with a randomized, spatially non-uniform distribution along the extended direction. 
This produces spatially inhomogeneous evaporation flows that lead to condensations forming self-consistently within the 3D magnetic structure. 
As a result, the model naturally develops a complex, thread-like morphology of cool plasma (shown by the temperature contours).
This illustrates that the evaporation--condensation mechanism remains viable in 3D geometry and produces prominence-like structures that are more comparable to observations.

%Using a similar dip-free magnetic arcade, \citet{fang2013,fang2015} extended the work of \citet{xiac2012} by exploring long-term coronal rain evolution with higher resolution and extended run times.
%In the following work, the role of localized footpoint heating in dip-free arcades has been further explored using different temporal structures. 
%\citet{lix2022} implemented a spatially and temporally stochastic heating model, finding that such heating configurations not only more readily trigger condensation, but also produce fragmented, short-lived, multiscale structures closely resembling observed coronal rain. 
%In follow-up studies \citep{lix2023, lix2025}, they examined the influence of flux emergence, demonstrating that reconnection events can significantly reshape condensation distributions and trigger dual-temperature jets.

More recently, \citet{zhou2023} and \citet{jerc2024} focused on arcade systems with pre-existing dips similar to \citet{kepp2014}. 
\citet{zhou2023} introduced periodic footpoint heating and demonstrated that the repeated energy input not only triggers condensation but also exerts an upward force on weakly magnetized regions, resulting in periodic vertical oscillations of the prominence. 
This lifting mechanism is rooted in magnetic pressure modulation caused by localized heating and was shown to operate even in the absence of external perturbations. 
Synthetic H$\alpha$ spectra derived from their model exhibited alternating intensity shifts between the red and blue wings—a pattern characteristic of the so-called ``winking filament” phenomenon\footnote{A winking filament refers to a filament that periodically brightens and darkens in different wings of the H$\alpha$ line, often interpreted as a signature of vertical oscillations along the line of sight, see, for example, \citet{hyde1966, rams1966, eto2002, okam2004}}, which has been reported in both ground-based and space-based observations. 
Their simulations successfully reproduced typical winking filament periods and velocity amplitudes, establishing a direct physical link between evaporation–condensation dynamics and this long-standing observational feature.
\citet{jerc2024} systematically compared steady and stochastic heating scenarios, showing that steady heating favors vertically extended, coherent filament structures, while stochastic heating induces fragmented, horizontally distributed threads with complex substructure and oscillatory dynamics. 
This highlights the sensitivity of prominence morphology to the temporal and spatial nature of chromospheric energy release.

Most recently, \citet{john2025} presented a 2.5D MHD simulation centered on a similar arcade configuration, but with a coronal X-point. 
Their model included evaporation--condensation processes, allowing condensations to accumulate above the null point and eventually force magnetic reconnection. 
This resulted in the partial drainage of filament plasma along newly reconnected field lines, forming a hybrid filament/coronal rain structure. 
Although distinct in topology, the simulation reinforced the role of mass–field coupling and chromospheric feedback in regulating condensation behavior.

\subsubsection{Road to Higher Dimensions and Higher Resolution}
While most 2D evaporation--condensation simulations are based on magnetic cross-sections, an alternative approach extends directly from the 1D flux-tube framework.  
Building on this idea, \citet{zhou2020} developed a flux sheet model composed of an array of field-aligned tubes, enabling limited transverse MHD coupling while preserving high ($\sim$10~km) spatial resolution along individual field lines, as shown by the cartoon in Fig.~\ref{fig7}(b).
Their simulations, incorporating radiative cooling, anisotropic conduction, and partial ionization effects, successfully reproduced spontaneous formation of thread-like condensations with observed-like widths ($\sim$100 km), lengths (20–30 Mm), and filling factors (10–15\%). 
The resulting synthetic H$\alpha$ (see Fig.~\ref{fig7}(c))and EUV images revealed that these threads aligned tightly with magnetic field lines (misalignment $<2^\circ$), validating the use of fine threads as magnetic tracers. 
Additionally, the model reproduced multi-temperature counterstreaming flows: cold plasma oscillates along threads ($\sim$10–25 km/s) while hot plasma flows oppositely in the inter-thread regions ($\sim$80–100 km/s), providing a unified picture for counterflow dynamics.
Building upon the same configuration, \citet{jerc2023} conducted a parameter survey examining the impact of heating intensity and height on condensation morphology. 
Their results confirmed that stronger localized heating accelerates thread formation and increases density, while higher heating locations influence onset timing and flow structures. 
These studies collectively highlight the importance of localized heating properties in shaping condensation dynamics.

A complementary approach to achieving high-resolution 3D prominence modeling is the pseudo-3D method, in which a large number of 1D simulations are projected onto magnetic field lines embedded in a 3D structure. 
This strategy retains high spatial fidelity along each tube while enabling three-dimensional visualization. 
\citet{luna2012} applied this approach to a sheared magnetic arcade using 125 field-aligned 1D simulations. 
Their results revealed two distinct populations of condensations: long-lived, mass-accumulating threads located in magnetic dips, and short-lived, rapidly draining blobs. 
The simulated thread lengths and spatial distribution, along with synthetic H$\alpha$ and EUV diagnostics, matched observations well. 
In a similar pseudo-3D framework, \citet{guoj2022} conducted simulations using a twisted magnetic flux rope configuration rather than a sheared arcade, see Fig.~\ref{fig7}(d). 
Their simulations—built upon the evaporation--condensation framework—also produced thread-like structures consistent with observational morphology, particularly in the presence of strong magnetic twist.

Another promising strategy for modeling multidimensional condensation involves simplifying the MHD equations under the rigid-field line approximation. 
This assumption allows one to reduce the full MHD system to a HD model with constrained motion along field lines—commonly referred to as the ``frozen-field hydrodynamic" (ffHD) approach.
Originally used in modeling coronal loops and flare heating \citep{moky2005, moky2008}, this framework has recently been extended to evaporation--condensation simulations. 
\citet{zhou2025} optimized the ffHD model for efficiency by introducing hyperbolic thermal conduction and the TRAC method. 
These enhancements enable effective condensation simulations at realistic resolution in 3D. 
Their application to a full 3D magnetic flux rope configuration produced a slab-like filament with realistic dimensions and internal thermodynamic properties, as shown in Fig.~\ref{fig7}(e).
The prominence formed exhibited key observational signatures, such as spine-aligned structure, low temperatures ($\sim$10$^4$ K), together with elevated densities in the expected range for prominence plasma. 
Forward-modeled H$\alpha$ and EUV emission maps further validated its agreement with observed filament morphology, see Fig.~\ref{fig7}(f).

All of the approaches discussed above—pseudo-3D magnetic flux tube mapping or ffHD—rely on approximations that decouple or constrain full magnetic dynamics. 
While these methods offer high resolution and computational efficiency, they do not capture the self-consistent evolution of the magnetic field during prominence formation. 
In contrast, \citet{xiac2016} presented the first true 3D MHD simulation of evaporation--condensation-driven prominence formation within a dynamically evolving magnetic flux rope.
Starting from a sheared arcade that developed into a flux rope via surface shearing and flux cancellation, they imposed localized footpoint heating to initiate chromospheric evaporation. 
Condensations naturally formed in the resulting magnetic dips, giving rise to a prominence composed of fragmented threads and blobs. 
The prominence exhibited realistic physical properties, including densities ($\sim$$1.9 \times 10^{10}$ cm$^{-3}$), temperatures ($\sim$17,500 K), and persistent vertical flows ($\sim$5–6 km/s). 
Notably, falling condensations deformed the magnetic field by dragging dips downward, demonstrating active mass–field coupling. 
A dynamic equilibrium emerged between evaporation-driven condensation and gravitational drainage. 
Synthetic EUV and H$\alpha$ images showed excellent agreement with observed filament and cavity morphology, firmly establishing the viability of in situ condensation in fully 3D magnetic environments.

\subsection{In-situ Condensation Models}
Recent years have seen a notable rise in simulations focusing on in-situ condensation, a category of prominence formation mechanisms distinct from evaporation-driven models.
It is crucial to recognize that in numerical simulations, condensation can often emerge under seemingly spontaneous conditions. 
Such behavior necessitates caution: simulations must distinguish physically grounded condensations from numerical artifacts. 
Reproducibility across grid resolutions, the influence of numerical schemes, and the identification of a coherent physical trigger are essential criteria for validating condensation events.
\subsubsection{Levitation--Condensation Model}
\begin{figure}[h]
\centering
\includegraphics[width=0.9\textwidth]{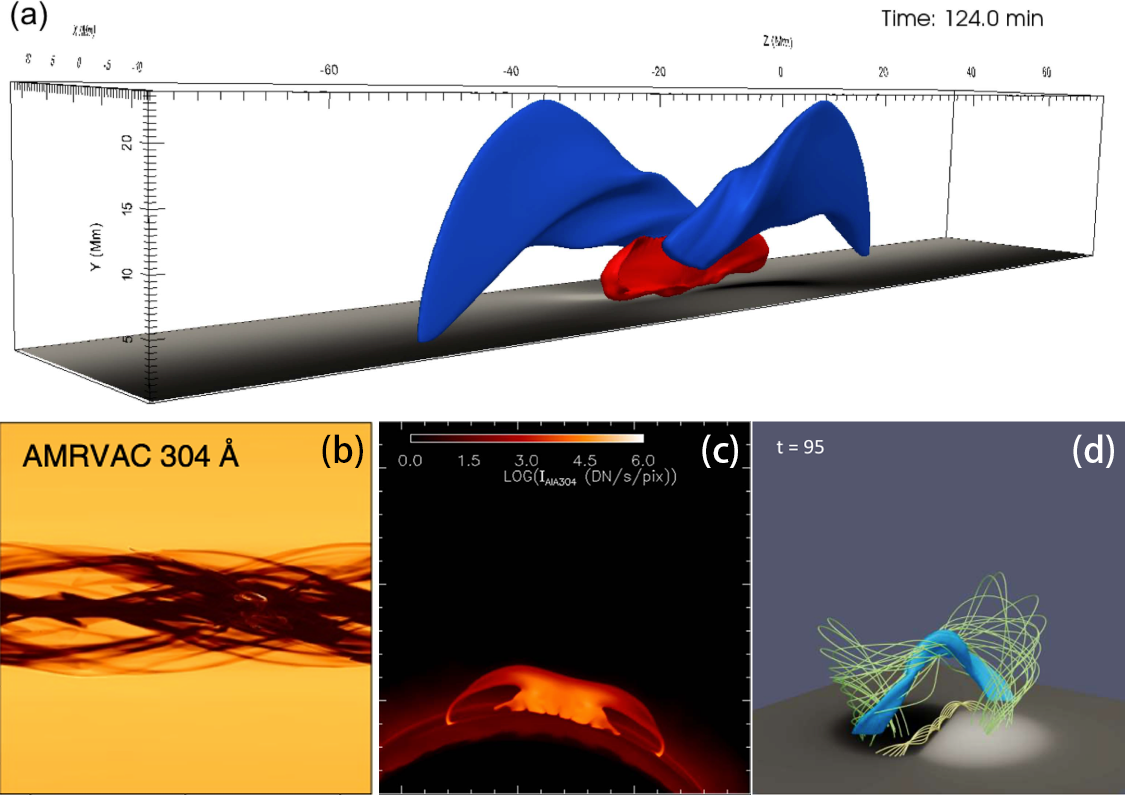}
\caption{
Examples of numerical simulations illustrating in-situ condensation and magnetic levitation.  
(a) 3D MHD simulation by \citet{kane2017}, showing the formation of a prominence via in-situ condensation.  
Red and blue isocontours represent cool plasma at different density thresholds embedded within the magnetic structure.  
(b) Synthetic EUV 304~\AA{} emission map from a high-resolution 3D simulation by \citet{jenk2022}, based on the 2D models of \citet{kane2015} and \citet{jenk2021}.  
The image shows the morphology of fine thread-like structure.
(c) EUV 304~\AA{} synthetic emission from the eruption simulation of \citet{fany2018a}, in which prominence material forms in situ during the flux rope eruption.  
(d) Snapshot from \citet{xing2025}, showing prominence mass (blue isocontours) being lifted from the lower atmosphere as part of a flux rope during its slow-rise phase, consistent with the levitation scenario.
}
\label{fig8}
\end{figure}

As mentioned, one representative framework for in-situ condensation is the so-called ``reconnection--condensation" or ``levitation--condensation" mechanism. 
\citet{kane2015, kane2017, kane2018} systematically developed this levitation--condensation model, proposing that magnetic reconnection can locally disrupt thermal equilibrium in the corona, triggering condensation.
In their 2.5D MHD simulations \citep{kane2015}, arcade field lines near PILs underwent reconnection due to converging and shear flows, producing a closed magnetic flux rope. 
Within this dense, low-lying coronal region, suppressed thermal conduction allowed radiative losses to dominate, leading to spontaneous condensation without any chromospheric mass input. 
This model offered a compelling explanation for observations lacking clear upward mass transport.
Subsequent work extended this framework to 3D \citep{kane2017}, incorporating anisotropic thermal conduction and full reconnection dynamics. 
They showed that the onset of condensation correlates with the field length, offering a criterion for instability.
Their results is shown in Fig.~\ref{fig8}(a).
Then, \citet{kane2018} introduced stochastic footpoint perturbations along the PIL, which led to spontaneous internal fragmentation, downflows, and plumes within the magnetic flux rope. 
A dynamic balance was established wherein condensation and gravitational drainage maintained a self-regulating mass cycle. 
This framework highlights the role of internal magnetic restructuring and perturbations in prominence formation, contrasting with chromospheric evaporation models.

Building on the model of \citet{kane2015}, \citet{jenk2021, jenk2022} advanced the levitation--condensation mechanism with higher-resolution simulations and full 3D MHD extensions. 
Their 2D AMR simulations \citep{jenk2021} reproduced magnetic flux rope formation and in-situ condensation with resolutions as fine as 6 km, revealing that condensation predominantly occurs in regions unstable to the magnetically projected Brunt--V\"ais\"al\"a criterion. 
Condensations formed at magnetic dips, then descended under gravity into equilibrium positions, producing monolithic suspended structures. 
The evolution involved vortical motions, filamentary thread formation, and quasi-periodic mass redistribution---highlighting the tight coupling between thermodynamics and field restructuring.
In the 3D extension, \citet{jenk2022} presented the first fully self-consistent 3D in-situ condensation model, forming a magnetic flux rope and subsequent condensation entirely within the coronal domain, without chromospheric coupling. 
Condensations initially formed at random locations and later coalesced into dips, accumulating $\sim$7$\times$10$^{12}$~g in mass over $\sim$15 minutes. 
Importantly, the study demonstrated the role of magnetic Rayleigh--Taylor instability (mRTI) in forming internal plume-like vertical structures. 
Synthetic EUV and H$\alpha$ emissions closely matched observations, capturing spine-thread and cavity structures, as shown in Fig.~\ref{fig8}(b).

Following \citet{jenk2021}, \citet{brug2022} tested various heating models to explore how background heating affects in-situ condensation. 
They introduced a reduced heating model based on magnetic geometry, which promoted earlier, more spatially extended condensation compared to traditional exponential heating.
The resulting condensations featured greater mass, more threads, and prominent non-isobaric evolution. 
The study showed that heating geometry critically influences both the onset and morphology of condensations.
Extending this framework further, \citet{donn2024} developed a coronal mass circulation model based on fully self-consistent 3D MHD, excluding chromospheric layers and artificial heating. 
Magnetic reconnection self-consistently formed a flux rope and triggered in-situ condensation, along with simultaneous coronal rain. 
By analyzing over 1000 field lines, they found that condensation occurs primarily above the flux rope in thermally unstable, overcooled pockets, with large-scale coronal siphon flows driven by TI-induced pressure gradients delivering mass.

\subsubsection{Other In-situ Condensation Models}
Another early example of in-situ condensation driven by magnetic activity is the work of \citet{choe1992}, who performed 2.5D magnetohydrodynamic simulations to investigate prominence formation triggered by photospheric shearing motions.  
In this class of models, condensation is not initiated by imposed artificial evaporation, but rather emerges as a consequence of coronal responses to magnetic field evolution---in particular, local density enhancements induced by magnetic expansion.
%Therefore, similar with the levitation--condensation model, but in a more general sense.
Specifically, their simulations showed that shear at the footpoints of a coronal arcade leads to upward expansion of magnetic field lines, which lifts denser plasma from lower altitudes into the upper corona.  
This process increases radiative losses in the loop apex region, triggering TI and the onset of localized condensation.  
The resulting structure develops into a vertically elongated, sheet-like prominence supported by dips in the reconfigured magnetic field.  
Mass accumulation was found to arise from both local condensation and siphon-like upflows, and the prominence eventually enters a quasi-dynamic state where new condensations form above and drain downward, producing a surrounding cavity of reduced density and pressure.

More recently, a series of 3D simulations by \citet{fany2017, fany2018a, fany2018b, fany2020} demonstrated in-situ condensation during the flux emergence process.
An snapshot from \citet{fany2018a} is shown in Fig.~\ref{fig8}(c).
Although primarily focused on eruption dynamics, these models showed that twisted flux ropes emerging from the solar interior naturally develop dipped field regions where TI triggers condensation, forming cool dense material ($T<10^5$~K) without chromospheric input. 
The PROM vs. non-PROM comparison in \citet{fany2020} confirmed that turning off TNE physics suppresses condensation, underscoring the role of TI and magnetic dips.

%Lastly, several post-eruption cooling studies \citep[e.g.,][]{ruan2021, sens2024} have examined coronal rain formation in flare-affected environments. 
%However, these have yet to robustly demonstrate in-situ prominence formation and are thus omitted from this review.

\subsection{Alternative Models: Injection and Levitation Models}\label{sec53}
From a numerical perspective, the injection and levitation models remain relatively underexplored compared to the extensively studied evaporation--condensation and in-situ condensation mechanisms. 
This lack of attention suggests that these alternative formation pathways for prominences may offer valuable opportunities for future investigation.

The injection model was first proposed in the 2D MHD simulations by \citet{anc1988}, where chromospheric material is directly supplied into the corona to form prominence structures. 
However, this approach received limited follow-up, possibly due to its apparent simplicity and intuitive nature. 
It is important to note that the physical parameters adopted in early simulations, constrained by computational resources at the time, were not fully representative of coronal conditions, limiting the realism and applicability of the results. 

More recently, the injection mechanism was revisited by \citet{huan2021}, who described a ``unified model.”
In their view, both evaporation–driven and injection–driven pathways could be realized within the same framework.
However, the “evaporation” part is debatable.
The imposed heating acts more like a perturbation to the corona, rather than the classical evaporation--condensation process described in Sect.~\ref{sec33}.

By contrast, the injection pathway is more straightforward.
In the lower atmosphere, impulsive energy release can drive an artificial jet that carries cool plasma upward into coronal structures.
This demonstrates that the injection mechanism can indeed produce prominence plasma.
However, the parameter space for effective injection is narrow:
{insufficient momentum fails to lift plasma into dips, while excessive momentum causes overshooting and fallback into the opposite footpoint.}
Although some observations suggest that cool plasma may indeed drain from the opposite side after injection, thereby circumventing the need for symmetric entrapment, this remains to be validated by future simulations.

{It is also worth noting that such jets are unlikely to be purely cold.
As discussed in Sect.~\ref{sec31}, solar jets often contain both hot and cool components.
In \citet{huan2021}, this mixed nature is hinted at but not very clear.
A more convincing demonstration was provided by \citet{lix2025}, who performed high-resolution 2D simulations of prominence formation triggered by magnetic flux emergence.
In their setup, emerging magnetic flux interacted with pre-existing coronal fields and led to reconnection.
The reconnection jets contained both hot and cool plasma components, and, importantly, they produced a prominence structure even without pre-existing magnetic dips.
This result shows that injection can operate self-consistently in realistic magnetic environments and that mixed hot–cold jets may provide an efficient channel for prominence mass supply.
A similar result has also been reported in the recent work of \citet{huan2025}.}

Recent progress has also been made in the development of the levitation model. 
A representative example is the 3D full-MHD simulation by \citet{xing2025}, which investigates the role of filament mass in the initiation of CMEs.  
The simulation employs shearing and converging surface flows to build a magnetic flux rope above the polarity inversion line, during which cool, dense plasma is gradually lifted into the corona, as shown in Fig.~\ref{fig8}(d).  
At first glance, the formation process resembles the levitation scenario, with magnetic reconnection producing dipped field lines that support filament mass.
However, the simulation setup closely parallels that used in earlier studies such as \citet{fany2017, fany2018a}, where in-situ condensation along coronal field lines has been shown to occur. 
{In addition, these simulations consistently reveal substantial drainage of cool plasma along the rising field lines.
This behavior not only provides an observationally testable signature of the levitation process, but also highlights a key limitation: the mass supplied purely by magnetic lifting is often insufficient to account for the observed prominence reservoir.}
It is therefore likely that a portion of the filament mass in this model also arises from localized radiative cooling. 
{In this sense, drainage during levitation may naturally create the conditions for subsequent in-situ condensation, further emphasizing the complementary roles of different mechanisms.}
Therefore, although the overall morphology of the filament in the \citet{xing2025} simulation is consistent with the lifting mechanism, the relative contributions from magnetic levitation and in-situ condensation remain difficult to quantify and require further investigation (private communication).

These developments suggest that while the injection and levitation models are less frequently employed, they hold considerable promise for enriching our understanding of prominence formation—especially in contexts where chromospheric connectivity is crucial or where mass loading alters eruption dynamics.

\section{Summary and Open Questions}\label{sec6}
Solar prominences, as dense and cool structures in the hot corona, continue to pose fundamental theoretical and observational challenges. 
In this review, we have surveyed the current understanding of prominence formation, with a focus on the physical origin of prominence plasma. 
We classified existing models into four representative categories: injection, levitation, evaporation--condensation, and in-situ condensation. 
These mechanisms were discussed in the context of their physical principles, numerical realizations, and observational implications.

Among the various formation mechanisms reviewed, in-situ condensation models were historically among the first to be proposed and remain observationally accessible, particularly in the context of coronal rain and low-lying prominences. 
In-situ condensation frequently appears in numerical simulations, often without deliberate triggering. 
However, the in-situ framework alone is not self-sufficient: it must explain what kinds of disturbances and under what conditions they can induce condensation in the corona. 
For example, the levitation--reconnection model posits that atmospheric disturbances caused by magnetic reconnection during the levitation process are responsible for initiating TNE and TI. 
Yet, the precise conditions---e.g., the nature of levitation, changes in field topology, or density enhancement---that enable such favorable environments remain to be quantified.

From this perspective, the evaporation--condensation model offers a more prescriptive mechanism: chromospheric heating induces upflows, increasing coronal density into a TNE regime. 
Once TI sets in, condensation ensues. 
Unlike in-situ condensation, the evaporation--condensation model explicitly identifies the chromosphere as the mass source, thus providing a natural explanation for the so-called "FIP problem" and prominence mass origin. 
The evaporation--condensation model has undergone extensive theoretical development in the past decade, particularly through high-fidelity simulations. 
However, observational confirmation remains sparse. 
One reason is that the evaporation phase requires high-resolution temporal and spectral diagnostics to detect. 
A central open question in this mechanism is the physical nature of the required localized heating---are nanoflares, chromospheric turbulence, or other mechanisms responsible? 
While quiescent regions may be compatible with evaporation--condensation formation, {this is because their relatively stable magnetic environment and long-lived heating conditions are more favorable for gradual evaporation--condensation cycles. 
In contrast, active regions are more likely to host impulsive, reconnection-driven processes consistent with the injection scenario.} 
The model faces challenges when applied to active regions or large-scale prominences with apexes above 50~Mm, where sustaining evaporation to such heights remains computationally difficult.
{This difficulty arises not only from numerical aspects---long-duration, high-resolution simulations are required to simultaneously capture the transition region, chromosphere, and extended coronal loops---but also from physical constraints. 
In particular, maintaining upflows over tens of Mm requires finely tuned, localized heating that is strong enough to drive evaporation yet not so strong as to disrupt the loop stability, 
while the large mass reservoir needed for quiescent prominences poses an additional challenge.}

The injection model, though conceptually simple, has gained support from recent observations, particularly of active region prominences. 
It is less likely to explain quiescent prominences due to their higher altitude. 
Injection models have not been widely explored in simulations, and like evaporation--condensation, they must account for the origin of the injection process. 
Preliminary results suggest that forming a stable, long-lived prominence via injection alone is nontrivial and requires fine-tuned parameters. 
However, the inherently dynamic nature of active region prominences---with frequent flows and mass exchange---suggests that a dynamically balanced state maintained by intermittent injections may be a more realistic and observationally consistent scenario.

Levitation models have received limited attention since their original proposal. 
Their robustness remains unclear, both observationally and in simulations. 
Notably, if levitation leads to atmospheric disturbances, it may naturally produce in-situ condensation---as encapsulated in the levitation--condensation framework. 
Even if the lifting mechanism is successful, it is likely to coincide with or induce other processes, making it difficult to isolate levitation as a standalone mechanism. 
Whether levitation should be regarded as an independent category thus remains debatable.

Although we have classified prominence formation models into four representative types, this division is primarily a theoretical convenience. 
In reality, these mechanisms are not mutually exclusive. 
As illustrated in the model by \citet{xing2025}, magnetic flux ropes may lift pre-existing mass (levitation), induce atmospheric disturbances that trigger in-situ condensation, and this may even incorporate injection from chromospheric jets. 
Especially in dynamic environments such as active regions, it is plausible that multiple mechanisms operate simultaneously. 
The classification serves as a framework for systematic study, but the actual formation of prominences in the solar corona likely results from a complex interplay of processes rather than a single dominant pathway.

To summarize, across mechanisms, several unifying themes have emerged. 
TI, while often central, is not sufficient alone to guarantee condensation; its interaction with magnetic topology, energy transport, and boundary conditions is critical. 
Magnetic dips serve as natural attractors for condensed plasma, regardless of how the mass is introduced. 
The spatial and temporal structure of heating---especially footpoint-localized heating---plays a key role in shaping the condensation morphology. 
{Additionally}, realistic prominence mass cycles must account for both mass accumulation and loss, and recent models have begun to reproduce self-sustained condensation-drainage equilibria.

Despite these advances, several outstanding questions remain:
\begin{itemize}
  \item \textbf{Trigger mechanisms:} What initiates condensation in realistic coronal conditions? How do small-scale perturbations such as reconnection, Alfv\'en wave dissipation, or magnetic flux emergence interact with global field structures to seed instability?
  \item \textbf{Chromospheric coupling:} Can prominence mass be fully sustained without chromospheric input? How does the inclusion of a resolved, more {realistic} chromosphere alter the dynamics of condensation and support?
  \item \textbf{Magnetic field diagnostics:} How can we constrain the supporting magnetic topology observationally? What synthetic diagnostics are needed to bridge the gap between MHD models and field measurements?
  \item \textbf{Mass budget and longevity:} What regulates the lifetime, mass content, and stability of prominences? 
  \item \textbf{Mechanism coexistence and transitions:} Under what conditions might one mechanism dominate or evolve into another?
\end{itemize}

Future progress will likely require further integration of high-resolution, multi-dimensional simulations with increasingly sophisticated observational diagnostics. 
The development of multi-scale models that couple chromospheric, coronal, and magnetic processes is essential. 
Next-generation instruments such as DKIST, EST and Solar-C will provide unprecedented opportunities for testing theoretical predictions. 
Finally, cross-disciplinary links with studies of coronal rain, solar wind, and CME initiation may offer new pathways for understanding the prominence mass problem in a broader heliophysical context.

\section*{Conflict of Interest}
The corresponding author states that there is no conflict of interest.

\bmhead{Acknowledgements}
YZ acknowledges funding from Research Foundation – Flanders FWO under project 1256423N, and support from the Jiangsu Provincial Double First-Class Initiative (grant number 1480604106).
%\bibliography{sn-bibliography}

\end{document}